# Counting methods introduced into the bibliometric research literature 1970 – 2018: A review

Marianne Gauffriau

Copenhagen University Library / The Royal Danish Library, Copenhagen, Denmark. E-mail: [mgau@kb.dk](mailto:mgau@kb.dk)

## Abstract

To design a bibliometric indicator, a counting method must be chosen. In many cases, this choice will affect the score obtained when applying the indicator. Thus, the study of counting methods is essential to the understanding of bibliometric indicators and to the informed choice of these indicators in practice.

The present review of counting methods investigates 1) the number of unique counting methods in the bibliometric research literature, 2) to what extent the counting methods can be categorized according to selected characteristics of the counting methods, 3) methods and elements to assess the internal validity of the counting methods, and 4) to what extent and with which characteristics the counting methods are used in research evaluations.

The review identifies 32 counting methods introduced during the period 1981 – 2018. Two frameworks categorize these counting methods. Framework 1 describes selected mathematical properties of counting methods, and Framework 2 describes arguments for choosing a counting method. Twenty of the 32 counting methods are rank-dependent, fractionalized, and introduced to measure contribution, participation, etc. of an object of study. Next, three criteria for internal validity are used to identify five methods that test the adequacy of counting methods, two elements that test sensitivity, and three elements that test homogeneity of the counting methods. These methods and elements may be used to assess the internal validity of counting methods. Finally, a literature search finds research evaluations that use the counting methods. Only three of the 32 counting methods are used by four research evaluations or more. Of these three counting methods, two are used with the same characteristics as defined in the studies that introduced the counting methods.

The review provides practitioners in research evaluation and researchers in bibliometrics with a detailed foundation for working with counting methods. At the same time, many of the findings in the review provide bases for future investigations of counting methods.

# 1 Introduction

The topic of the present review is counting methods in bibliometrics. The bibliometric research literature has discussed counting methods for at least 40 years. However, the topic remains relevant as the findings in the review show. Section 1 provides background for the review (Section 1.1), followed by the study aims (Section 1.2) and research questions (Section 1.3).

## 1.1 Background

The use of counting methods in the bibliometric research literature is often reduced to the choice between full and fractional counting. Full counting gives the authors of a publication 1 credit each, while fractional counting shares 1 credit between the authors of a publication. However, several studies document that there are not just two, but many counting methods in the bibliometric research literature, and that the distinction between full and fractional counting is too simple to cover these many counting methods (for examples, see Todeschini & Baccini, 2016, pp. 54–74; Waltman, 2016, pp. 378–380; Xu et al., 2016).

A counting method functions as one of the core elements of a bibliometric indicator, for example, the *h*-index fractionally counted (Egghe, 2008) or by first-author counting (Hu et al., 2010). For some indicators, the counting method is the only element, for example, the number of publications (full counting) on a researcher's publication list. Counting methods are not only relevant for how to construct and understand indicators but also for bibliometric network analyses (Perianes-Rodriguez et al., 2016), field-normalization of indicators (Waltman & van Eck, 2015), and rankings (Centre for Science and Technology Studies, n.d.).

There are bibliometric analyses where the choice of counting method makes no difference. For sole-authored publications the result is the same whether the authors are credited by full counting (1 credit) or fractional counting (1/1 credit). However, co-authorship is the norm in most research fields, and the average number of co-authors per publication has been increasing (Henriksen, 2016; Lindsey, 1980, p. 152; Price, 1986, pp. 78–79; Wuchty et al., 2007). Other objects of study reflect this trend; for example, the number of countries per publication has also been increasing (Gauffriau et al., 2008, p. 152; Henriksen, 2016).

Choosing a counting method is essential for the majority of bibliometric analyses that evaluate authors, institutions, countries, or other objects of study. A study of a sample of 99 bibliometric studies finds that, for two thirds of the studies, the choice of counting method can affect the results (Gauffriau, 2017, p. 678). The effect of shifting between counting methods can be seen in the *Leiden Ranking*, where it is possible to choose between full and fractional counting for the indicators on scientific impact (Centre for Science and Technology Studies, 2019). A change from one counting method to the other alters the scores, and thus, the order of the institutions in the ranking.

Nonetheless, many bibliometric studies do not explicitly justify the choice of counting method. More broadly, in bibliometric practice, there are no common standards for how to describe counting methods in the methods section of studies (Gauffriau, 2017, p. 678; Gauffriau et al., 2008, pp. 166–169). This implicit status of counting methods is also reflected in bibliometric textbooks and handbooks from the most recent decade. Many do not have a chapter or larger section dedicated to counting methods (for examples, see Ball, 2018; Cronin & Sugimoto, 2014; Gingras, 2016; W. Glänzel et al., 2019). Others have a chapter or larger section dedicated to counting methods but include only common counting methods and / or lack well-defined frameworks to describe the counting methods (for examples, see Sugimoto & Larivière, 2018, pp. 54–56; Todeschini & Baccini, 2016, pp. 54–74; Vinkler, 2010, Chapter 10; Vitanov, 2016, pp. 27–29).

The present review demonstrates that a consistent analysis of the majority of bibliometric counting methods can reveal new knowledge about those counting methods. Given this, the discussion argues for the explicit use, analysis, and discussion of counting methods in bibliometric practice and research as well as in bibliometric textbooks and handbooks.

## 1.2   Aims and relevance

The topic of this review is bibliometric counting methods. The aims are to investigate counting methods in the bibliometric research literature and to provide insights into their common characteristics, the assessment of their internal validity, and how they are used. The review presents different categorizations of counting methods and discusses the counting methods based on these categorizations. Hence, the review does not focus on counting methods individually but rather on the general characteristics found across counting methods. The general characteristics provide insights into the counting methods and how they overlap or differ from each other.

Three previous reviews of counting methods (Todeschini & Baccini, 2016, pp. 54–74; Waltman, 2016, pp. 378–380; Xu et al., 2016) are fairly comprehensive; however, the present review includes still more counting methods. In their handbook, Todeschini & Baccini present a list of counting methods and provide a definition for each counting method, but they do not in a consistent manner analyze characteristics across counting methods. Waltman uses a division into full, fractional, and other counting methods. The present review develops Waltman's approach further, resulting in a more well-defined categorization of counting methods. Xu et al.'s categorization of counting methods analyzes data distributions. Although the review does not use this categorization, it discusses the categorization as one route for future research.

Counting methods are often categorized with regard to their mathematical properties (for examples, see Rousseau et al., 2018, sec. 5.6.3; Waltman, 2016, pp. 378–380; Xu et al., 2016). In addition to a categorization based on selected mathematical properties (Gauffriau et al., 2007), the review applies another approach (Gauffriau, 2017), which builds on qualitative text analysis. The review adopts this approach to describe why counting methods are introduced into the bibliometric research literature.

Previous studies either have documented that many different counting methods exist in the bibliometric research literature or have analyzed selected counting methods using well-defined frameworks. The present review does both by covering more counting methods than previous reviews and by providing detailed insight into the general characteristics of these counting methods. Furthermore, the review considers three criteria for assessing the internal validity of bibliometric indicators (Gingras, 2014), applying these to identify methods and elements that can be used to assess the internal validity of the counting methods. Finally, the review investigates the use of the counting methods in research evaluations. The results of the review are a unique resource for informing the use of counting methods and inspiring further investigations of counting methods.

## 1.3   Research questions: RQs 1 – 4

The aims presented in Section 1.2 lead to four inter-connected research questions (RQs):

>   RQ 1: How many unique counting methods are there and when were they introduced into the bibliometric research literature?

RQ 1 is useful to understand the magnitude and timeliness of this review's aims. As discussed in Section 1.1, counting methods often remain implicit in bibliometric analyses, even though there are many counting methods to choose from and a change from one counting method to another may alter the scores for the objects of study. The review provides an overview of how many counting methods there are in the bibliometric research literature. Making this information available is the first step in facilitating the explicit and informed choice of counting methods in bibliometric analyses.

> RQ 2: To what extent can the counting methods identified by RQ 1 be categorized according to selected frameworks that focus on characteristics of the counting methods?

RQ 2 explores whether the counting methods identified by RQ 1 share characteristics. As Section 1.1 mentions, the simplified dichotomy full or fractional counting is often seen in the bibliometric research literature. The analysis for RQ 2 uses two fine-grained frameworks to provide both a more detailed categorization of the counting methods' mathematical properties as well as a categorization of why the counting methods were introduced into the bibliometric research literature. These categorizations do not focus on only a few counting methods, but rather, provide knowledge about a large number of counting methods.

> RQ 3: Which methods and elements from the studies that introduce the counting methods identified by RQ 1 can be used to assess the internal validity of those counting methods?

Where RQ 2 focuses on shared characteristics of the counting methods identified by RQ 1, RQ 3 supplements this with information about the assessment of the internal validity of counting methods, i.e. as drawn from the studies that introduce the counting methods. As discussed in Section 1.1, the counting method is a core element in the construction of a bibliometric indicator. Therefore, not only the characteristics of the counting methods but also the internal validity of the counting methods are important.

> RQ 4: To what extent are the counting methods identified by RQ 1 used in research evaluations and to what extent is this use compliant with the definitions in the studies that introduce the counting methods?

As mentioned previously, the use of counting methods is often reduced to a choice between full and fractional counting. RQ 4 investigates the use of counting methods in more detail. The use of the counting methods should comply with the design of the counting methods. If one or more of the characteristics identified under RQ 2 change between the point of introduction and the point of use of a counting method then the internal validity of the counting method may be compromised.

# 2   Methods

Section 2 presents the methods used to address the RQs presented in Section 1.3. Organized by RQ, Table 1 summarizes the methods, as well as the related data, tools, and results. Sections 2.1 – 2.4 provide detailed presentations of the methods, discuss the rationale for applying the methods, and show how the results from each RQ inform the subsequent RQs.

Table 1: Summary of RQs and their research methods, data, tools, and results.

| RQ | Method | Data | Tools | Results |
|---|---|---|---|---|
| RQ 1 | Literature search based on citing and cited studies. | Peer-reviewed studies in English published 1970 – 2018. | *Google Scholar.* | Thirty-two counting methods introduced into the bibliometric research literature over the period 1981 – 2018. No unique counting methods are introduced in the period 1970 – 1980. |
| RQ 2 | Categorizations of counting methods. | The 32 counting methods identified by RQ 1. | Two frameworks: The first describes selected mathematical properties of counting methods, and the second describes arguments for choosing a counting method. | Thirty counting methods are categorized according to the first framework, and all 32 counting methods are categorized according to the second framework. |
| RQ 3 | Identification of methods and elements useful for assessing internal validity of counting methods. | The 32 counting methods identified by RQ 1. | Three internal validity criteria for bibliometric indicators. The results from RQ 2. | Five methods and five elements related to the assessment of the internal validity of counting methods are identified. |
| RQ 4 | Literature search based on citing studies. | The 32 counting methods identified by RQ 1. A sample of research evaluations that use counting methods identified by RQ 1. | *Google Scholar*. The results from RQ 2. | Three of the 32 counting methods are each used in a minimum of four research evaluations. For one counting method, the use does not comply with the characteristics of the counting method as defined by the study that introduced the counting method. |

## 2.1 RQ 1: Literature search for counting methods

RQ 1 serves to illustrate the magnitude and timeliness of the present review. The results of RQ 1 go on to form the basis for RQs 2 – 4.

> RQ 1: How many unique counting methods are there and when were they introduced into the bibliometric research literature?

To answer RQ 1, a literature search is employed to identify counting methods in the bibliometric research literature. The literature search concentrates on studies that introduce counting methods rather than studies that use counting methods.

To be included in the review, studies must introduce counting methods defined by an equation or similar to guide calculation. Where a study presents only minor variations to the equation of an existing counting method, that variant approach is not included as a separate counting method. Section 3.2 gives a few examples of such variations. Variations of existing counting methods are also not included in cases where the variations add weights to publication counts in the form of citations, *Journal Impact Factors*, publication type weights, etc. Furthermore, the counting methods must be applicable to publications with any number of authors. For counting methods introduced with different objects of study in different studies — for example, institutions in one study and countries in another study — only the first study is included.

In addition to the counting methods included in this review, 'hybrids' also exist in the literature. Hybrids are counting methods that sit somewhere between two known counting methods, for example, Combined Credit Allocation that sits between complete-fractionalized and Harmonic counting (X. Z. Liu & Fang, 2012, p. 41) or First and Others credit-assignment schema that sits between straight and Harmonic counting (Weigang, 2017, p. 187). Hybrids are not included in the review.

The literature search is restricted to peer-reviewed studies in English from the period 1970 – 2018. Prior to 1970, discussions about counting methods in bibliometrics seem to be in the context of specific studies; however, from approximately 1970 onwards, some of the discussions about counting methods start to offer general recommendations in relation to choosing a counting method for a bibliometric analysis. These general recommendations derive primarily from changing norms for co-authorship and the launch of the *Science Citation Index* (Cole & Cole, 1973, pp. 32–33; Lindsey, 1980; Narin, 1976), but other factors such as institutional and national research policies and programs may have had an effect as well. Counting methods introduced after 2018 are not included in the review as the use of these is difficult to assess (RQ 4) at the present time, i.e. less than two years after their introduction into the bibliometric research literature.

Studies that introduce counting methods into the bibliometric research literature are found via a search of cited and citing studies (Harter, 1986, pp. 58–60; 186). The search begins with studies cited by or citing Gauffriau et al.'s work on counting methods (Gauffriau, 2017; Gauffriau et al., 2007, 2008; Gauffriau & Larsen, 2005). From these cited and citing studies, studies are selected for the review. The cited and citing studies of these selected studies are then searched to add yet more studies to the review, and so on. This search approach is chosen because the terminology for counting methods is not well defined. Many terms are general, for example, 'count' or 'number of', making adequate keywords difficult to identify.

Citations to ten selected studies are searched using *Google Scholar*, *Scopus*, and *Web of Science*. *Google Scholar* proves to have the best coverage. This finding is supported by large scale studies of *Google Scholar* (Delgado López-Cózar et al., 2019, sec. 4.3).

Two sources are used to find cited and citing studies, respectively. Cited studies are found via the reference lists in the selected studies. Citing studies are found via *Google Scholar*. Titles and the first lines of the abstracts as presented in *Google Scholar's* list of results are skimmed to find relevant studies among the citing studies. In addition, a search in the citing studies for the terms 'count', 'counting', 'fraction', 'fractionalized', and 'fractionalised' is conducted via *Google Scholar's* 'Search within citing articles'. The results are skimmed to find relevant studies. The final search for citing studies is completed in December 2018.

The result of the literature search cannot claim a 100 percent coverage of all counting methods that exist in the bibliometric research literature; however, when citations to and from the studies included in this review are checked, a significant redundancy is encountered, i.e. studies already included in the review. Thus, it is assumed that the review covers the majority of counting methods discussed in the bibliometric research literature during the period 1970 – 2018.

## 2.2   RQ 2: Categorizations of counting methods according to their characteristics

RQ 1 identifies counting methods in the bibliometric research literature. RQ 2 explores whether the counting methods identified under RQ 1 share characteristics. Shared characteristics can facilitate general knowledge about counting methods rather than specific knowledge of only a few counting methods. The results of RQ 2 are used in the analyses related to RQs 3 and 4.

> RQ 2: To what extent can the counting methods identified by RQ 1 be categorized according to selected frameworks that focus on characteristics of the counting methods?

The list of 32 counting methods created under RQ 1 does not provide information about the types of counting methods. Therefore, RQ 2 applies two frameworks to categorize the counting methods identified by RQ 1. 'Framework' is a unifying term for generalizable approaches that describe and facilitate the use, analysis, and discussion of counting methods. These approaches compile elements such as consistent terminology, definitions, and categorizations of counting methods.

The first framework (Framework 1) describes selected mathematical properties of the counting methods (Gauffriau et al., 2007). The second framework (Framework 2) is based on a qualitative text analysis and describes four groups of arguments for the choice of counting method in a bibliometric analysis (Gauffriau, 2017). As such, the two frameworks have different foundations and they address different characteristics of the counting methods. The frameworks are described in detail in Sections 2.2.1 and 2.2.2.

### 2.2.1   Framework 1: Selected mathematical properties of counting methods

'Framework 1: Selected mathematical properties of counting methods' builds on measure theory (Halmos, 1950) and provides well-defined definitions and a detailed terminology for counting methods (Gauffriau et al., 2007). Thus, the framework offers a more precise terminology compared to the dichotomy full and fractional counting.

The review uses the framework to categorize counting methods according to two mathematical properties: Whether a counting method is rank-dependent or rank-independent, and whether it is fractionalized or non-fractionalized.

Below, firstly, the framework's terminology for counting methods is introduced (Gauffriau et al., 2007). Secondly, the mathematical properties of the counting methods are presented (ibid.). And lastly, assumptions made about the mathematical properties are discussed to enable the use of the framework in the present review.

**Terminology**

Counting method

A "counting method is defined by the choice of basic unit [of analysis], object [of study] and score function" (Gauffriau et al., 2007, p. 178; square brackets added). The Danish Bibliometric Research Indicator (Agency for Science and Higher Education, 2019, sec. Co-authored publications) can serve as an illustration of the triplet: basic units of analysis, objects of study, and score function. For simplification, the collaboration bonus and publication type weights in the Danish Bibliometric Research Indicator are not included in the example that follows: A publication has three authors and two institutions. Author A and author B are from institution Y, and author C is from institution Z. According to the Danish Bibliometric Research Indicator, authors are the basic units of analysis. The authors are credited via a score function, which in the Danish Bibliometric Research Indicator is fractional counting. Institutions, the objects of study, obtain scores by collecting the credits from their affiliated authors. Thus, the two institutions' scores are Y = 1/3+1/3 = 2/3 and Z = 1/3.

The present review restricts the literature search to unique score functions rather than unique counting methods introduced into the bibliometric research literature. One score function can be applied in several counting methods. That is, score functions can be used with different combinations of basic units of analysis and objects of study. This means that a score function can describe a class of counting methods. However, from the literature search, it is clear that score functions are introduced as counting methods with specific basic units of analysis and objects of study, often authors (i.e. at the micro-level). Therefore, the term counting method is used for the score functions identified in the literature search.

Objects of study and basic units of analysis

Objects of study are the objects presented in the results of bibliometric analyses, for example, researchers, institutions, or countries. Objects of study can be found in publications, but they may also be objects not directly visible in the publication, for example, unions of countries (e.g. the European Union (EU) and the United Kingdom).

Basic units of analysis are found in publications.

Objects of study are scored by collecting credits from the basic units of analysis. It is common to find bibliometric analyses in which the basic units of analysis and the objects of study are at the same aggregation level, for example, authors (micro-level). However, in the Danish Bibliometric Research Indicator, the basic units of analysis are authors (micro-level) and the objects of study are institutions (meso-level). Another example is the calculation of the score for the EU as an object of study. Credits are given to countries belonging to the EU; thus, countries are the basic units of analysis. Thus, the objects of study and the basic units of analysis are at different aggregation levels: unions of countries (supra-level) and countries (macro-level), respectively.

Score function

A score function describes how the objects of study are scored. The basic units of analysis are credited individually before the objects of study collect the credits. Five common score functions are presented below.

The five score functions:

- Complete
  A credit of 1 is given to each basic unit of analysis in a publication. An object of study collects the credits from the basic units of analysis assigned to the object of study.

- Complete-fractionalized
  A credit of $1/n$ is given to each basic unit of analysis where $n$ is the number of basic units of analysis in a publication. An object of study collects the credits from the basic units of analysis assigned to the object of study.

- Straight
  A credit of 1 is given to the basic unit of analysis ranked first in a publication. All other basic units of analysis in the publication are credited 0. An object of study collects the credits from the basic units of analysis assigned to the object of study.

  Instead of first authors (i.e. the basic unit of analysis ranked first in the publication), last authors or reprint authors can also be credited (for examples, see Gauffriau et al., 2007, p. 676). The review does not discuss these alternatives further.

- Whole
  A credit of 1 is given to each basic unit of analysis, assigned one-to-one to a unique object of study, in a publication. If a unique object of study is represented by more basic units of analysis, in a publication, these basic units of analysis share 1 credit in whatever way. An object of study collects the credits from the basic units of analysis assigned to the object of study.

- Whole-fractionalized
  A credit of $1/m$ is given to each basic unit of analysis, assigned one-to-one to a unique object of study, where $m$ is the number of unique objects of study related to a publication. If a unique object of study is represented by more basic units of analysis, in a publication, these basic units of analysis share $1/m$ credit in whatever way. An object of study collects the credits from the basic units of analysis assigned to the object of study.

When the terminology is reduced to full and fractional counting, the difference between complete and whole score functions are not immediately visible. Both are called full counting. Neither are the differences between complete-fractionalized, straight, and whole-fractionalized score functions. All three are variations of fractional counting.

A note on terminology

In measure theory, which is the theoretical basis for Framework 1, the term 'normalized' is used (Halmos, 1950, p. 171) for the property where the credit of 1 is shared, i.e. divided amongst the basic units of analysis of a publication. The present review uses the alternative term 'fractionalized' because this has become the norm in the bibliometric research literature discussing counting methods. The term 'normalized' typically refers to field-normalization (Waltman, 2016, secs. 6 and 7).

**Mathematical properties**

The five score functions above have definitions based on five mathematical properties introduced below. Table 2 shows how the mathematical properties form score functions, and thus, classes of counting methods (Gauffriau et al., 2007, p. 198). Detailed explanations follow below the table. The five mathematical properties are used to form assumptions, which are necessary for the analysis undertaken in relation to RQ 2.

Table 2: Decision tree for the different score functions and classes of counting methods.[1]

| Defined for all objects | Based on a fixed crediting scheme | Additive | Rank-independent | Fractiona-lized | Classes of counting methods described in the literature |
|---|---|---|---|---|---|
| Yes | Yes | | Yes | No | Complete |
| | | | | Yes | Complete-fractionalized |
| | | | No | No | |
| | | | | Yes | Straight |
| | No | Yes | Not applicable | No | |
| | | | | Yes | |
| | | No | | No | Whole |
| | | | | Yes | |
| No | No | No | Not applicable | No | |
| | | | | Yes | Whole-fractionalized |

The five mathematical properties:

- Defined for all objects / not defined for all objects
  All classes of counting methods in Table 2 except whole-fractionalized counting are defined for all objects of study. To test whether a counting method is defined for all objects of study, some of the objects of study can be merged to form a union. If this does not change the score for the objects of study not included in the union, then the score function is defined for all objects of study. The UK as object of study can be used as illustration. To find all publications affiliated with the UK in *Web of Science*, it is necessary to search for publications from England, Scotland, Wales, and Northern Ireland. Take a publication with ten unique country affiliations in which the UK is represented by only one of England, Scotland, Wales, or Northern Ireland. Following whole-fractionalized counting, the score for each country affiliated with the publication is 1/10. Now take another publication, again with ten unique country affiliations. In this publication, the UK is represented by three countries, for example, England, Scotland, and Wales. In this case, the three countries are merged, and the score for each country affiliated with the publication becomes 1/8.

---

[1] (Gauffriau et al., 2007, p. 189). The term normalized is changed to fractionalized and the column "In Section" is removed.

- Based on a fixed crediting scheme / not based on a fixed crediting scheme
  All classes of counting methods in Table 2 except whole and whole-fractionalized counting have fixed crediting schemes. Whole and whole-fractionalized counting are not based on fixed crediting schemes, as a change of objects of study may also change the credits given to basic units of analysis. If the objects of study and the basic units of analysis are institutions, then unique institutions in the affiliation section of a publication will be credited. If the basic units of analysis are kept and the objects of studies changed to countries, then unique countries in a publication will be credited via their institutions. If more than one institution from a country contributes to the publication, then the institutions share the credit for that country. Thus, the basic units of analysis cannot be credited independently of the objects of study. If a counting method is based on a fixed crediting scheme, then the counting method is additive (see next item).

- Additive / non-additive
  Complete, complete-fractionalized, and straight counting are additive. The score for the objects of study can be calculated via credits to basic units of analysis at the same aggregation level (for example, macro-level) or to basic units of analysis at lower aggregation levels (for example, meso- or micro-level) and the score will remain the same given that there is a one-to-one relation between the aggregation levels. In other words, if countries are objects of study, then it makes no difference whether the basic units of analysis are institutions or countries, providing that each address in the affiliation section of a publication has only one institution and one country. If a counting method is additive, then the counting method is defined for all objects (see first item).

- Rank-independent / rank-dependent
  Complete and complete-fractionalized counting are rank-independent. The order of the basic units of analysis — for example, the order of countries in the affiliation section of a publication — does not influence how the basic units of analysis are credited. All basic units of analysis get the same credit. Straight counting is rank-dependent because only the first basic unit in the affiliation section of a publication is credited. All other basic units of analysis get 0 credit. This property of rank-independency / rank-dependency is not applicable to whole and whole-fractionalized counting as these counting methods are not based on fixed crediting schemes. For example, if countries are the objects of study, then for a publication with ten country affiliations, where affiliation number two, six and seven are Denmark, the credit can be attributed to the affiliation ranked second, sixth, or seventh in whatever way. Thus, rank-dependency cannot be applied. Neither can rank-independency be applied, i.e. where all basic units of analysis receive the same credit.

- Fractionalized / non-fractionalized
  Complete-fractionalized, straight, and whole-fractionalized counting are fractionalized because, with these methods, the basic units of analysis in a publication share a total credit of 1. The rationale is that a publication equals 1 credit. Complete and whole counting are not fractionalized, i.e. as the credits for the basic units of analysis of a publication can sum to more than 1. Note that fractionalized and additive are two different properties. For example, whole-fractionalized counting is fractionalized and non-additive, whereas complete counting is non-fractionalized and additive.

In the review, the use of the framework with these five mathematical properties to categorize counting methods incorporate the assumptions below.

**Assumptions about mathematical properties for counting methods**

The analyses presented in the review focus on two of the five properties: rank-independent / rank-dependent and fractionalized / non-fractionalized. The following assumptions explain why the review focuses on these two properties.

As already mentioned, score functions are introduced into the bibliometric research literature as counting methods, often with authors as basic units of analysis and objects of study (i.e. at the micro-level). Without information about how the score functions work at, for example, the meso- or macro-level, it is difficult to decide the score functions' status for the first three mathematical properties: defined for all objects / not defined for all objects, based on a fixed crediting scheme / not based on a fixed crediting scheme, and additive / non-additive. For example, complete-fractionalized and whole-fractionalized counting differ regarding the three properties (see Table 2), but at the micro-level, the calculations of scores are identical. At other aggregation levels, the calculations differ for the two counting methods.

Using Framework 1, however, the first three mathematical properties can help making assumptions about the counting methods at the micro-level that use rank to determine credits for the basic units of analysis. As mentioned in the introduction to the five mathematical properties, counting methods with rank-dependent score functions have a fixed crediting scheme. If based on a fixed crediting scheme, the score functions are additive. If additive, the score functions are defined for all objects.

For score functions introduced as counting methods at the micro-level that are not rank-dependent, it is difficult to decide if the score function is rank-independent (for example, complete and complete-fractionalized counting) or, rather, if the rank-independent / rank-dependent property is not applicable (for example, whole and whole-fractionalized counting). In the review, such counting methods are assumed to be rank-independent, and thus, based on a fixed crediting scheme, additive, and defined for all objects.

For all counting methods included in the review the status for the property fractionalized / non-fractionalized is explicitly evident in the studies that introduce the counting methods.

Based on the above assumptions, the results of the present review focus on the properties rank-dependent / rank-independent and fractionalized / non-fractionalized. Thus, the categorization of counting methods is: rank-dependent and fractionalized (see Section 3.2.1), rank-dependent and non-fractionalized (see Section 3.2.2), rank-independent and non-fractionalized (see Section 3.2.3), and rank-independent and fractionalized (see Section 3.2.4).

### 2.2.2 Framework 2: Four groups of arguments for choosing a counting method for a study

'Framework 2: Four groups of arguments for choosing a counting method for a study' proposes a categorization of arguments for choosing a counting method for a study. The categorization is developed from the arguments for counting methods in a sample of 32 studies published in 2016 in peer-reviewed journals and supplemented with arguments for counting methods from three older studies (Gauffriau, 2017).

The review uses Framework 2 to categorize counting methods according to the arguments for why a counting method is introduced into the bibliometric research literature. The studies found in relation to RQ 1 that introduce counting methods argue for why the new counting methods are needed. These arguments are assigned to the four groups of arguments in Framework 2. This use is a slight modification compared to the original intention of Framework 2, in which the arguments relate to choosing a counting method for a

study – not introducing a new counting method. However, the review assumes that a counting method is introduced with the aim of being used in other studies. Thus, the argument for the introduction and for the use of a counting method are seen as compatible.

Table 3 presents the categorization with Groups 1 – 4 (Gauffriau, 2017, p. 679). Descriptions of the four groups follow Table 3.

Table 3: Categorization of arguments for counting methods for publication and citation indicators.[2]

| Category | Counting method(s) |
|---|---|
| **Group 1: The indicator measures the (impact of)…** | |
| … participation of an object of study | Whole |
| … production of an object of study | Whole, complete-fractionalized |
| … contribution of an object of study | Whole, complete-fractionalized (rank-independent and rank-dependent) |
| … output / volume / creditable to / performance of an object of study | Whole, complete-fractionalized |
| … the role of authors affiliated with an object of study | Straight, last author, reprint author |
| **Group 2: Additivity of counting method** | |
| Additivity of counting method | Whole, complete-fractionalized |
| **Group 3: Pragmatic reasons** | |
| Availability of data | Whole, straight, reprint author |
| Prevalence of counting method | Whole |
| Simplification of indicator | Whole |
| Insensitive to change of counting method | Whole |
| **Group 4: Influence on / from the research community** | |
| Incentive against collaboration | Complete-fractionalized |
| Comply with researchers' perceptions of how their publications and / or citations are counted | Whole |

The four groups of arguments for choosing a counting method:

- Group 1: The indicator measures the (impact of) contribution / participation / … of an object of study
  The arguments for counting methods relate to the concept that the study attempts to measure by using the counting method to design an indicator. For example, some studies in the sample argue that whole counting is suitable for indicators measuring the objects of study's participation in a research endeavor.

- Group 2: Additivity of counting method
  The arguments for counting methods relate to mathematical properties of the counting method itself: namely, to ensure that the counting method is additive and to avoid double counting of publications.

- Group 3: Pragmatic reasons
  The conceptual / methodological arguments included in Groups 1 and 2 are not taken into account but instead are pragmatic reasons for the choice of a counting method. Whole counting is quite common in

---
[2] (Gauffriau, 2017, p. 679)

Group 3. This may be explained by this counting method being the readily available approach in the databases often used to calculate bibliometric indicators (i.e. *Web of Science* and *Scopus*). In these databases, a search for publications from Denmark returns the number of publications in which Denmark appears at least once in the list of affiliations. This corresponds to whole counting.

- Group 4: Influence on / from the research community
  The arguments in Group 4 are not related to what an indicator measures (i.e. as in Group 1), but rather, to the impact of the indicator on the research community under evaluation (and vice versa). For example, one of the studies analyzed to create the framework argued for whole counting when the objects of study are researchers because a researcher should have 1 credit for each publication in his / her publication list (Waltman & van Eck, 2015, p. 891). The argument is that this is how a researcher intuitively counts his / her publications, and that this intuitive counting approach should be reflected in the evaluation.

In the review, the categorization of counting methods uses all four groups of arguments. The analysis focuses on arguments for why the counting methods are introduced into the bibliometric research literature.

## 2.3 RQ 3: Internal validity of counting methods

RQ 2 focuses on shared characteristics of the counting methods identified by RQ 1. RQ 3 adds information about the assessment of internal validity of the counting methods by the studies that introduce the counting methods.

> RQ 3: Which methods and elements from the studies that introduce the counting methods identified by RQ 1 can be used to assess the internal validity of those counting methods?

To answer RQ 3, methods for and elements of the assessment of internal validity of the counting methods in the studies that introduce the counting methods (RQ 1) are identified. There are no standards commonly applied for such assessments of internal validity, and only a few of the studies that introduce counting methods explicitly include assessments of the internal validity of the counting methods. However, all the studies include analyses of the counting methods. These analyses set out methods that may be used to assess the internal validity of the counting methods. As well, the counting methods may have elements that themselves can indicate weak internal validity of the counting methods.

It is not possible to evaluate in a consistent and manageable manner how well these methods and elements work as assessments of internal validity of counting methods. Instead, RQ 3 evaluates how well each of the methods and elements corresponds to three criteria for well-constructed bibliometric indicators: Adequacy, sensitivity, and homogeneity (Gingras, 2014, pp. 112–116). In the review, the criteria are used on counting methods instead of indicators. Counting methods, however, function as core elements or the only element in bibliometric indicators. Thus, if a counting method does not comply with the criteria for internal validity, the same conclusion could be reached for bibliometric indicators using that counting method.

Guidance is provided for how to apply the criteria at an overarching level (ibid.). However, implementation in a specific case, such as the present review, requires several choices as described below. Apart from the study introducing the three criteria, two studies have applied the three criteria to evaluate bibliometric indicators (Wildgaard, 2015a, sec. 6.3, 2019, sec. 14.4.1). The present review is the first to use to three criteria to evaluate counting methods.

Three validity criteria for well-constructed bibliometric indicators:

- Adequacy
  According to the adequacy criterion, an indicator should be an adequate proxy for the object the indicator is designed to measure. The indicator and the object should have the same characteristics, for example, order of magnitude. The relationship between object and indicator is tested via an independent and accepted measure for the object (Gingras, 2014, pp. 112–115).

  The counting methods identified by RQ 1 are assigned, using Framework 2, to arguments for the introduction of the counting methods. In the implementation of the adequacy criterion, the methods below may be used to document that the counting methods are adequate proxies for their aims, that is, the arguments for the counting methods:
    o Compare to other counting methods or bibliometric indicators: The scores obtained by the counting method are compared to scores obtained by existing counting methods or bibliometric indicators when applied on empirical publication sets or publication sets constructed for exemplification. Some publication sets are as small as one publication.
    o Principles to guide the definitions of the counting methods: A list of principles are stated explicitly and used in the definition of the counting method.
    o Quantitative models for distributions of scores: A quantitative model is used to test whether the counting method gives scores, fitting the model, to the objects of study.
    o Surveys or other empirical evidence: Surveys or other empirical evidence about co-authorship practice are used to define target values for the credits for basic units of analysis.
    o Compare groups of objects of study: The scores obtained by the counting method are compared for groups of objects of study with different characteristics.

  There are more methods for the assessment of the adequacy of a counting method, but each of these methods was found only in one study and, therefore, not included in the list above. One example is the comparison of scores obtained by the counting method where the order of the authors in a publication is kept versus where the order is shuffled (Trueba & Guerrero, 2004, fig. 4).

  As mentioned, the present review's analysis does not assess how well these methods work in the studies that introduce counting methods. Instead, the analysis assesses whether the methods are appropriate to test the counting methods as adequate proxies for their aims.

- Sensitivity
  According to the sensitivity criterion, an indicator should reflect changes over time in the object that the indicator is designed to measure (Gingras, 2014, pp. 115–116).

  Section 1 explains that the increasing average number of co-authors per publication is a driver behind the discussion about counting methods. In relation to the sensitivity criterion, two elements are defined. Where present, these elements highlight counting methods that are less flexible to an increasing number of authors per publication:

- o Time specific evidence: Surveys or other empirical evidence about co-authorship practice that are not updated over time and, therefore, do not reflect changes over time in the average number of co-authors per publication.
- o Fixed credits for selected basic units of analysis: A fixed share of the credit for selected basic unit of analysis, for example, the first author. As the number of authors per publication increases, such fixed credits leave less credit left for the non-selected authors of the publications. This is only true for fractionalized counting methods where 1 credit is shared among the authors of a publications. Therefore, the analysis considers this element in relation to fractionalized counting methods only.

Counting methods with one or both of the above elements are less flexible to an increasing number of authors per publication and, therefore, they do not comply with the sensitivity criterion.

- According to the homogeneity criterion, an indicator should measure only one dimension and avoid heterogeneous indicators, for example, the *h*-index, which combines publication and citation counts in one indicator. When a heterogeneous indicator increases / decreases, it is not immediately clear whether one or more elements cause the change. Thus, the indicator becomes difficult to interpret (Gingras, 2014, p. 116).

  Some of the counting methods are homogeneous, whereas others are complex, mixing many elements. A mix of elements can make it difficult to instantly understand how the scores of the counting method are obtained and which elements account for how much of the score. The implementation of the homogeneity criterion investigates elements that work against the criterion:
    - o Parameter values selected by bibliometrician: The equation for the counting method has parameter(s) where the bibliometrician selects the values of the parameter(s) for each analysis individually.
    - o External elements: The equation for the counting method is dependent on elements external to the publications that are included in an analysis (for example, an author's position as principal investigator, an author's *h*-index, an author's number of publications, etc.)
    - o Conditional equations: To calculate credits for all basic units of analysis, a conditional equation is needed. One part of the equation is dedicated to specific basic units of analysis or specific publications (for example, first authors or publications with local authors) and another part of the equation is dedicated to the remaining basic units of analysis or publications.

  Counting methods with one or more of the above elements are heterogeneous, as the elements lead to several dimensions being present in the same counting method.

  Of the three validity criteria, the homogeneity criterion is the most difficult to apply to counting methods. A mix of different elements that have the same measure unit does not count as heterogeneous but as composite (Gingras, 2014, p. 122). This said, the difference between heterogeneous and composite is described using an example (ibid.), which makes an exact interpretation difficult. It is a matter for debate whether some of the conditional equations included in the review's analysis use the same measure unit — for example, author contributions for first authors and other authors, respectively — and, therefore, whether these equations identify true heterogeneous counting methods.

## 2.4 RQ 4: Use of the counting methods in research evaluations

RQ 4 investigates to what extent the counting methods identified by RQ 1 are used in research evaluations. The research evaluations should comply with the design of the counting methods in the studies that introduce the counting methods. If one or more of the characteristics identified under RQ 2 change from the introduction to the use of the counting methods, then the introducing study's guidance about how to use the counting method may be compromised.

> RQ 4: To what extent are the counting methods identified by RQ 1 used in research evaluations and to what extent is this use compliant with the definitions in the studies that introduce the counting methods?

RQ 4 is addressed through a literature search aimed at identifying research evaluations that use the counting methods identified by RQ 1. The literature search is restricted to peer-reviewed studies in English. The peer-review criterion ensures some level of quality check and increases the likelihood that researchers have authored the studies. Thus, reports from university management, PowerPoint presentations, sales materials, etc. are not included. The literature search does not distinguish between studies where the research evaluations are the primary result and studies where the research evaluations are part of the results.

Counting methods can be used in many contexts, for example, in the development of new counting methods or investigations of the mathematical properties of the counting methods. The focus for the present review is research evaluations covering a minimum of 30 researchers where researchers are the objects of study. If institutions or countries are the objects of study, the institutions or countries cannot be represented by fewer than 30 researchers. Counting methods that are difficult to apply on larger publication sets are probably less well-suited for research evaluations. In other words, the emphasis is on scalable counting methods.

To find studies that use the counting methods identified by RQ 1, citations in *Google Scholar* to the counting methods are searched. As discussed in Section 2.1, *Google Scholar* covers more publications relevant to the review than either *Web of Science* or *Scopus*. Furthermore, to avoid research evaluations with almost identical implementations of a counting method, the same author cannot represent several research evaluations for the same counting method. Some counting methods are used in several studies by the same author (for example, see, Abramo et al., 2013, p. 201, 2020, p. 7). Including all of these research evaluations would give this implementation more weight than implementations represented by one research evaluation.

The number of research evaluations that use each counting method is reported using the following intervals: zero research evaluations use the counting method, one to three research evaluations use the counting method, and four or more research evaluations use the counting method.

For counting methods with four or more research evaluations, samples of five research evaluations, if available, are selected randomly for inclusion in an analysis of the use of the counting methods. With five research evaluations per counting method, it is possible to get an indication of whether or not the characteristics from the introduction of the counting methods, as identified under RQ 2, are kept in the research evaluations. With one to three research evaluations per counting method the results of the analysis would not be sufficiently robust.

## 2.5 Summary of methods

Section 2 presents the methods used to address RQs 1 – 4 and shows how each RQ informs subsequent RQs. The review uses a range of methods, starting with a literature search to identify counting methods. The results of this literature search go on to inform all remaining RQs. Next, two frameworks are used to categorize the counting methods. In addition, three validity criteria are used to identify methods and elements in the studies that introduce the counting methods to assess the internal validity of counting methods. Finally, a literature search finds research evaluations that use the counting methods. This subsequent use in practice of the counting methods is compared to the initial definition of the counting methods.

# 3 Results

Section 3 reports the results for RQs 1 – 4 based on the methods presented in Section 2. Sections 3.1 – 3.4 present the results for each of the RQs 1 – 4. Section 3.5 summarizes the results. The Supplementary Material (see Section 7.1) offers a schematic overview of results under all RQs.

## 3.1 RQ 1: Thirty-two unique counting methods in the bibliometric research literature

RQ 1: How many unique counting methods are there and when were they introduced into the bibliometric research literature?

Four score functions are introduced prior to 1970 and fall outside the time frame covered by the literature search. Recall that score functions are counting methods where different basic units of analysis and objects of study can be applied. The four score functions are complete, complete-fractionalized, straight, and whole counting (see definitions in Section 2.2.1). The review uses these pre-1970 score functions as reference points in some of the following analyses.

Beyond the four pre-1970 score functions, another 32 unique score functions are identified. These are introduced into the bibliometric research literature during the period 1981 – 2018.[3] There are no unique score functions introduced during the period 1970 – 1980. The majority, or 17, of the score functions are introduced in the most recent decade (2010 – 2018), as illustrated in Figure 1.

All the score functions are introduced as counting methods, which are score functions with specific units of analysis and objects of study. Thus, the term 'counting methods' is used for the score functions identified in the literature search.

---

[3] Two counting methods were published online first in 2018 and, thus, included in the period covered by the review. The two studies introducing the counting methods were assigned to journal issues in 2019 (Bihari & Tripathi, 2019; Steinbrüchel, 2019).

Figure 1: Number of unique counting methods introduced into the bibliometric research literature 1970 – 2018.

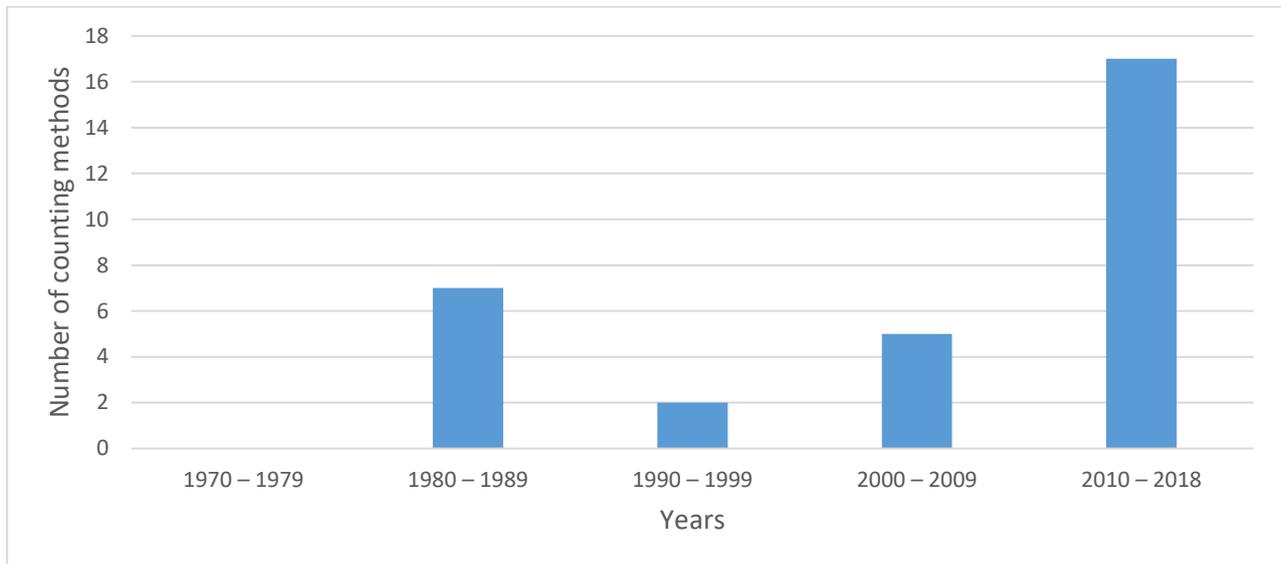

## 3.2  RQ 2: Categorizations of counting methods according to Frameworks 1 and 2

RQ 2: To what extent can the counting methods identified by RQ 1 be categorized according to selected frameworks that focus on characteristics of the counting methods?

The RQ 2 categorizations build on two frameworks. The frameworks are independent of each other. In the presentation of the results, Framework 1 with selected mathematical properties takes priority. This framework can be seen as a further development of the binary division into full and fractional counting – a division often seen in discussions about counting methods. Framework 2 describes arguments for choosing a counting method. With Framework 2 as secondary, the framework adds extra information to the categories created via Framework 1. However, it is possible for either of the frameworks to be given priority or for the categorizations of the counting methods to be presented separately for each framework. To support different categorizations, the Supplementary Material (Section 7.1) simply lists all 32 counting methods chronologically.

In the presentation below, beginning with the largest category, the counting methods are divided into four categories based on Framework 1: rank-dependent and fractionalized (Section 3.2.1), rank-dependent and non-fractionalized (Section 3.2.2), rank-independent and non-fractionalized (Section 3.2.3), and rank-independent and fractionalized (Section 3.2.4). Two counting methods do not fit these properties (Section 3.2.5) and two arguments for introducing counting methods do not currently comply with Framework 2 (Section 3.2.6).

Most of the counting methods have a name. Counting methods without a name are in the review named after the author(s) of the study introducing the counting methods, i.e. [author].

### 3.2.1 Rank-dependent and fractionalized counting methods

Twenty-one of the 32 counting methods identified by RQ 1 are rank-dependent and fractionalized meaning that the basic units of analysis in a publication share 1 credit but do not receive equal shares. Among the pre-1970 counting methods, straight counting has these properties.

In addition to the rank-dependent counting methods, the results include counting methods where the credits for basic units of analysis are shared unevenly based on characteristics other than rank, for example, an author's position as principal investigator, an author's $h$-index, or an author's number of publications.

Figure 2 and List 1 present the 21 counting methods. In Figure 2, a ten-author publication example illustrates the 14 counting methods[4] where rank determines the credits for the basic units of analysis. Figure 2 is followed by List 1, with seven counting methods where characteristics other than rank determine the credits. For the counting methods in List 1, more information than the number and rank of authors in a publication is needed to calculate the credits. This extra information is, for example, an author's position as principal investigator, an author's $h$-index, or an author's number of publications. Thus, it is not possible to do a generic calculation for a publication with ten authors and show the seven counting methods in Figure 1. Instead, List 1 describes these counting methods.

Eighteen of the 21 counting methods are defined with authors as both basic units of analysis and objects of study. One counting method is defined with authors as basic units of analysis and institutions as objects of study (Howard et al., 1987, p. 976). One study, which introduces two counting methods, has authors and countries as basic units of analysis and objects of study (Egghe et al., 2000, p. 146).

The arguments for 18 of the 21 counting methods can be linked to Group 1 in Framework 2: 'The indicator measures the (impact of) contribution / participation / … of an object of study'. In addition, two of the 21 counting methods (Assimakis & Adam, 2010; Howard et al., 1987, p. 976) aim to measure productivity,[5] an approach that is not included but can be added to Group 1 in Framework 2. The final study of the 21 studies argues "Credit is allocated among scientists based on their perceived contribution rather than their actual contribution" (Shen & Barabasi, 2014, p. 12329). This argument is assigned to Group 4 in Framework 2: 'Comply with researchers' perceptions of how their publications and / or citations are counted'.

---

[4] The counting method Weighted fractional output is in Figure 2 with its two versions: Intra-mural is used for publications where first and last author are from the same institution. Otherwise, extra-mural is used (Abramo et al., 2013, p. 201). In the counting method Network-based model, the bibliometrician must select a distribution factor (Kim & Diesner, 2014). Two distribution factors (d=0.25 and d=0.59) are selected for the illustration in Figure 2.

[5] The term 'productivity' is debated. Often, it is used as a simple concept, as is the case in the two referenced studies. This simple interpretation of productivity can be added to Group 1 in Framework 2. However, Abramo & D'Angelo argue for productivity as a complex concept requiring input and output indicators to calculate the productivity of the objects of study (Abramo & D'Angelo, 2014). Abramo & D'Angelo introduce a counting method relating to output (see Figure 2). The argument for their counting method is to measure author contributions (Abramo et al., 2013, p. 200). This argument is assigned to Group 1 in Framework 2: 'The indicator measures the (impact of) contribution / participation / … of an object of study'. Thus, in Abramo & D'Angelo's interpretation of productivity is the counting method one of the steps in calculating productivity (Abramo & D'Angelo, 2014, pp. 1135–1136).

Figure 2: How authors of a publication with ten authors share the credit. Rank-dependent and fractionalized counting methods.

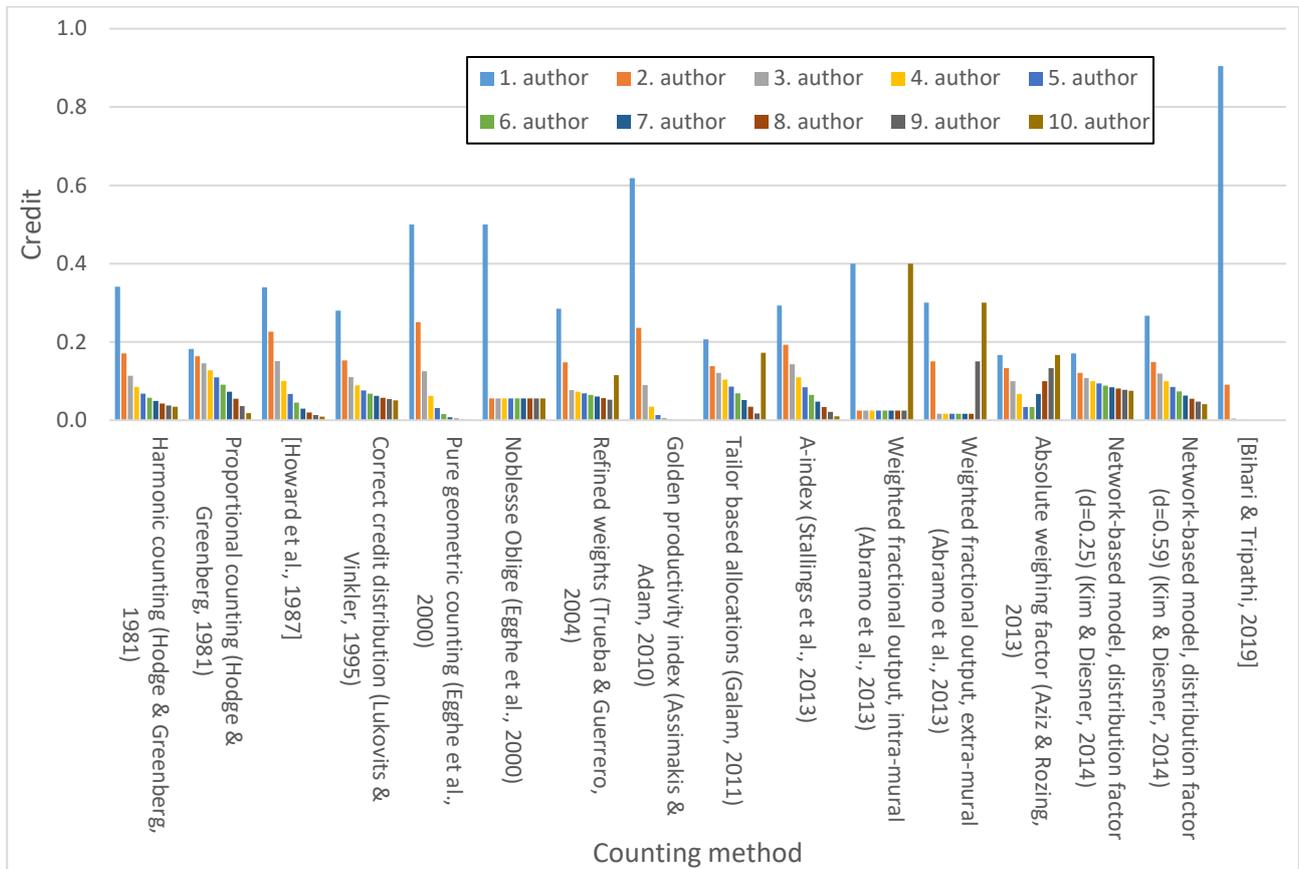

As mentioned above, in addition to the 14 counting methods comprising Figure 2, there are seven counting methods where the credits for basic units of analysis are shared unevenly based on characteristics other than rank. List 1 describes these counting methods.

All but one of the counting methods in List 1 were introduced after Framework 1 was published. In the framework, rank is determined based on the information in a publication (Gauffriau et al., 2007, pp. 179; 188). Thus, the definition of rank in the framework does not cover the counting methods where the credits are distributed based on characteristics other than rank. This review assumes credits distributed based on characteristics other than rank to be a variation of the property rank-dependent. However, the counting methods in List 1 are defined with authors as both basic units of analysis and objects of study. It would require information about how the counting methods are defined at other aggregation levels to find out whether or not credits distributed based on characteristics other than rank can be seen as a variation of the property rank-dependent for these counting methods.

In List 1, there are some examples of studies that add small changes to existing counting methods. As discussed in Section 2.1, the review does not consider these studies as presenting distinct representations of counting methods.

List 1: Fractionalized counting methods. The credits to the basic units of analysis are distributed based on characteristics other than rank.

- [Boxenbaum et al.]
  The credit of 1 for a publication is divided between the authors in such a way that the senior author receives twice the credit of non-senior authors (Boxenbaum et al., 1987, pp. 566–568).

- Pareto weights
  The credit of 1 for a publication is divided between the authors. An author receives the greater credit if the number of actual citations is more in line with the author's average number of citations per publication, i.e. neither higher nor lower (Tol, 2011, pp. 292–293; 296–297).

  Persson suggests a modification of Tol's counting method where the weight assigned to an author can change from one publication to the next (Persson, 2017). The review does not discuss these alternatives further.

- Shapley value approach
  The credit of 1 for a publication is divided between the authors according to their Shapley value. Shapley value is a concept from game theory. An author's weight is calculated by averaging the marginal contribution of the author in all possible co-author combinations. The marginal contribution is based on the author's number of citations, or on other impact scores for the author (Papapetrou et al., 2011).

- Contribution to an article's visibility, first approach
  The credit of 1 for a publication is divided between the authors, with weights dependent on an indicator (for example, the *h*-index) calculated for each author (Egghe et al., 2013, pp. 57–59).

- [Shen & Barabasi]
  The credit of 1 for a publication is divided between the authors, with weights dependent on the author's share of authorships in the co-citation network of the publication and also on the number of co-citations. The more publications and citations an author has in the research field, the more credit will be assigned to her / him (Shen & Barabasi, 2014).

  Other studies suggest modifications to Shen & Barabasi's counting method. In one study, author ranks in the publications are taken into account (Wang et al., 2017); in another, publication years and whether or not publications are highly cited are taken into account (Bao & Zhai, 2017). The review does not discuss these alternatives further.

- Relative intellectual contribution
  In publications where the authors state their contributions guided by the *CRediT*[6] taxonomy, the types of contributions can be weighted and these weights credited to the contributing authors. In total, all author contributions to a publication sum to 1 (Rahman et al., 2017).

---

[6] *CRediT* is a taxonomy for describing the contributions made by authors to research publications: https://casrai.org/credit/

- [Steinbrüchel]
  The credit of 1 for a publication is divided equally between those authors who are principal investigators. All other authors of the publication are credited 0 (Steinbrüchel, 2019, pp. 307–308).

### 3.2.2 Rank-dependent and non-fractionalized counting methods

A much smaller category, with six counting methods, has the properties rank-dependent and non-fractionalized, meaning that the sum of credits for basic units of analysis in a publication can sum to more than 1 credit and the basic units of analysis do not receive equal shares. The pre-1970 counting methods are not represented in this category.

In addition to the rank-dependent counting methods, the results include counting methods where the credits for basic units of analysis are shared unevenly based on characteristics other than rank, for example, an author's position as principal investigator, an author's *h*-index, or an author's number of publications.

The six counting methods are defined with authors as basic units of analysis and objects of study.

The arguments for the introductions of four of the six counting methods are from Group 1 in Framework 2: 'The indicator measures the (impact of) contribution / participation / … of an object of study'. The two remaining counting methods (Ellwein et al., 1989, p. 320) aim to measure productivity,[7] an approach that is not included in but can be added to Group 1 in Framework 2.

Figure 3 and List 2 present the six counting methods. Figure 3 shows five of the counting methods. An example with a ten-author publication provides a visual representation of the counting methods. For one of the counting methods, the credits are distributed based on characteristics other than rank as discussed in relation to List 1. List 2 with only one item describes this counting method.

Figure 3: How authors of a publication with ten authors share the credit. Rank-dependent and non-fractionalized counting methods.

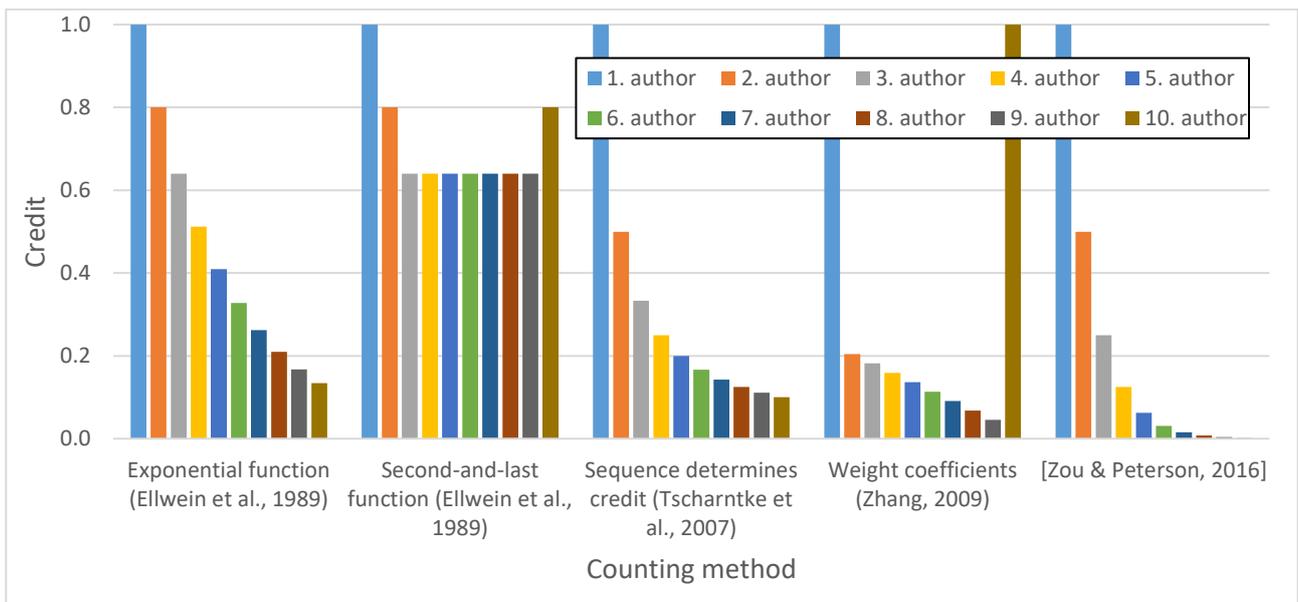

---

[7] Ellwein at al. (1989) use the simple interpretation of the term productivity. See Footnote 5.

List 2: Non-fractionalized counting method. Credits to the basic units of analysis are distributed based on characteristics other than rank.

- Contribution to an article's visibility, second approach
  The *h*-index (or another indicator) is calculated for each author of a publication and for the union of the authors' publications. An author receives a share of the credit for the publication equal to her/his *h*-index divided by the *h*-index for the union. The sum of credits to the authors of a publication may exceed 1 (Egghe et al., 2013, pp. 57–59).

### 3.2.3 Rank-independent and non-fractionalized counting methods

The next category includes three counting methods, which are rank-independent and non-fractionalized. All basic units of analysis in a publication receive equally-sized credits and the total credit for a publication can sum to more than 1. Among the pre-1970 counting methods, complete counting has these properties.

The three counting methods are defined with authors as basic units of analysis and objects of study.

The fist counting method aims to give a balanced representation of productivity across research disciplines (Kyvik, 1989, pp. 206–209). This type of argument is not yet included in Framework 2. See Appendix 1 for a further analysis. For the two remaining counting methods (de Mesnard, 2017; Tscharntke et al., 2007), the argument for the introduction of the counting method is assigned to Group 1 in Framework 2: 'The indicator measures the (impact of) contribution / participation / … of an object of study'.

Figure 4 provides a visual representation of the three counting methods. A ten-author publication is used as an example for the illustration. In Figure 4, the Equal Contribution method results in scores identical to scores obtained by complete-fractionalized counting. However, complete-fractionalized counting has no limit for how small a fraction of the credit, from a publication to a basic unit of analysis, can be. The Equal Contribution method gives minimum 5% of the credit from a publication to a basic unit of analysis; thus, unlike scores obtained by complete-fractionalized counting, the total credit for a publication can sum to more than 1.[8]

---

[8] In the bibliometric research literature, the Danish Bibliometric Research Indicator is described with different calculations (Nielsen, 2017, p. 3; Schneider, 2009, p. 372; Wien et al., 2017, pp. 905–907). Applying authors as basic units of analysis and objects of study, the calculation of the Danish Bibliometric Research Indicator overlap with the Equal Contribution method with the modification that 10% of the credit is the minimum credit from a publication to a basic unit of analysis. However, the Danish Bibliometric Research Indicator has authors as basic units of analysis and institutions as objects of study. Thus, the calculation of the Danish Bibliometric Research Indicator differs from the Equal Contribution method (see a further discussion in Section 3.2.6).

Figure 4: How authors of a publication with ten authors share the credit. Rank-independent and non-fractionalized counting methods.

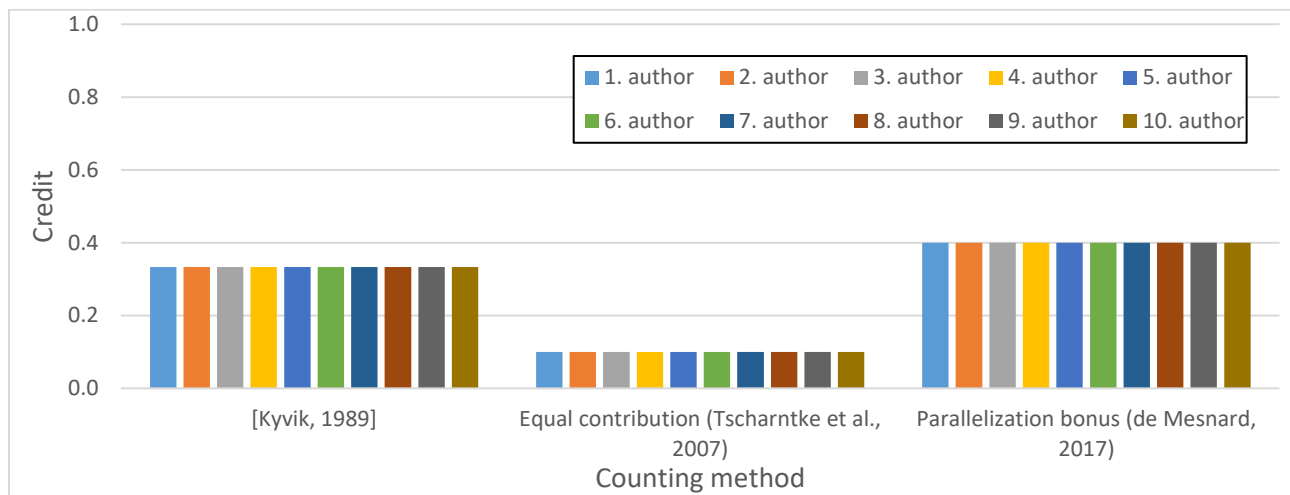

### 3.2.4 Rank-independent and fractionalized counting methods

For completeness, the category of rank-independent and fractionalized counting methods is included. However, none of the 32 counting methods identified by RQ 1 belong to this category. Among the pre-1970 counting methods, complete-fractionalized counting has the properties of being rank-independent and fractionalized. The basic units of analysis in a publication share 1 credit evenly.

### 3.2.5 Two counting methods do not comply with Framework 1

Two of the counting methods identified by RQ 1 do not comply with the selected properties from Framework 1: rank-dependent / rank-independent and fractionalized / non-fractionalized. These are the On-line fractionation approach (Nederhof & Moed, 1993) and the Norwegian Publication Indicator (NPI) (Sivertsen, 2016, p. 912). As documented below, there are different reasons for why the two counting methods do not fit Framework 1.

The two counting methods are analyzed under RQ 3 and 4. These analyses are not affected by the counting methods not fitting Framework 1.

**On-line fractionation approach**

The description of Framework 1 in Section 2.2.1 mentions that whole counting and whole-fractionalized counting do not comply with the property of being rank-dependent or rank-independent. Whole-fractionalized counting is introduced in 1993, under the name On-line fractionation approach (Nederhof & Moed, 1993). As such, it is well-documented that the On-line fractionation approach does not comply with the Framework 1 property of being either rank-dependent or rank-independent (Gauffriau et al., 2007, p. 188). Nonetheless, the basic units of analysis and objects of study can be determined. On-line fractionation approach is defined with countries as basic units of analysis and objects of study. The argument for the introduction of the counting method is assigned to Group 3 in Framework 2: 'Pragmatic reasons'. The counting method is easier to use on larger publication sets compared to complete-fractionalized counting (Nederhof & Moed, 1993, p. 41).

**The counting method used in the Norwegian Publication Indicator (NPI)**

The other counting method that does not comply with Framework 1 is the Norwegian Publication Indicator (NPI) (Sivertsen, 2016, p. 912). In the NPI, authors are basic units of analysis and institutions are objects of study. An institution's score for a publication is calculated by first adding up the complete-fractionalized credits of the authors from the institution to a sum for the institution. Next, the square root of the sum is calculated. Applying the square root to a sum for basic units of analysis as done in the NPI does not comply with measure theory (von Ins, personal communication, 30[th] Oct., 2018), which is the theoretical foundation for the mathematical properties of Framework 1 (see Section 2.2.1).[9]

Furthermore, neither does the NPI comply with Framework 2. Similar to Kyvik's counting method (Kyvik, 1989, pp. 206–209), the argument for the NPI is to give a balanced representation of productivity across research disciplines (Sivertsen, 2016, p. 912). This argument has not yet been covered by Framework 2.

The NPI does not fit with either of the Frameworks 1 and 2. Thus, Appendix 1 is a case study that uses the two frameworks to analyze the NPI, and through this analysis, identify potential for developing the frameworks further.

### 3.2.6  Counting methods that do not currently comply with Framework 2

Two arguments for introducing counting methods do not currently comply with Framework 2. Both arguments are reported in the sections above.

The first argument is to give a balanced representation of productivity across research disciplines used in two studies (Kyvik, 1989, pp. 206–209; Sivertsen, 2016, p. 912). This argument has not yet been covered by Framework 2; however, in Appendix 1, a case study discusses how Framework 2 can be developed to include the argument.

The second argument is to measure productivity, an approach that is not included but can be added to Group 1 in Framework 2 as mentioned in Sections 3.2.1 and 3.2.2. This argument is used by three studies (Assimakis & Adam, 2010, p. 422; Ellwein et al., 1989, p. 320; Howard et al., 1987, p. 976).

The counting methods are analyzed in relation to RQs 3 and 4. These analyses are not affected by the counting methods not currently fitting Framework 2.

---

[9] The Danish Bibliometric Research Indicator follows similar steps in the calculation. An institution's score from a publication is calculated by first adding up the complete-fractionalized credits of the authors from the institution to a sum for the institution. Next, institutions with less than 10% of the credit from a publication each have their credit raised to 10% of the credit from the publication (Agency for Science and Higher Education, 2019, sec. Fraktionering (in Danish)). This practice is discussed further for the NPI in Appendix 1 in Section 7.2.

## 3.3 RQ 3: Methods and elements to assess internal validity of counting methods

RQ 3: Which methods and elements from the studies that introduce the counting methods identified by RQ 1 can be used to assess the internal validity of those counting methods?

RQ 3 applies three criteria for well-constructed bibliometric indicators: Adequacy, sensitivity, and homogeneity. The adequacy criterion identifies methods that may be used to assess the adequacy of counting methods in the studies that introduce counting methods. The sensitivity and homogeneity criteria identify elements that indicate weak sensitivity and define heterogeneity in the equations of the counting methods and, as such, work against the two criteria.

RQ 3 does not evaluate how well these methods and elements work as assessments of internal validity of the counting methods in the studies that introduce counting methods. Thus, RQ 3 does not answer whether the 32 counting methods identified by RQ 1 are internally valid or not. Instead, RQ 3 evaluates how well each of the methods and elements corresponds to the relevant criteria for internal validity: Adequacy, sensitivity, and homogeneity.

Table 4 presents the schematic overview from the Supplementary Material (see Section 7.1) in relation to RQ 3. The table shows which counting methods that apply to the methods and elements to assess adequacy, sensitivity, and homogeneity. Sections 3.3.1 – 3.3.3 report the results for each of the three criteria.

Table 4: Overview of the 32 counting methods identified by RQ 1 in relation to the three criteria adequacy, sensitivity, and homogeneity.

| Counting method | Methods to support adequacy ||||| Elements that work against sensitivity || Elements that work against homogeneity |||
|---|---|---|---|---|---|---|---|---|---|---|
| | Compare to other counting methods or bibliometric indicators | Principles to guide the definitions of the counting methods | Quantitative models for distributions of scores | Surveys or other empirical evidence | Compare groups of objects of study | Time specific evidence | Fixed credits for selected basic units of analysis | Parameter values selected by bibliometrician | External elements | Conditional equations |
| Harmonic counting (Hodge & Greenberg, 1981; Hagen, 2008) | N | Y | N | N | N | N | N | N | N | N |
| Proportional counting (Hodge & Greenberg, 1981; Van Hooydonk, 1997) | N | Y | N | N | N | N | N | N | N | N |
| [Howard et al., 1987] | Y | N | N | N | Y | N | N | N | N | N |
| [Boxenbaum et al., 1987] | N | N | N | Y | N | Y | N | N | Y | Y |
| [Kyvik, 1989] | Y | N | Y | N | Y | N | NA | N | N | Y |
| Exponential function (Ellwein et al., 1989) | Y | N | N | N | Y | N | NA | Y | N | N |
| Second-and-last function (Ellwein et al., 1989) | Y | N | N | N | Y | N | NA | Y | N | N |
| On-line fractionation approach (Nederhof & Moed, 1993) | Y | N | N | N | Y | N | N | N | N | N |
| Correct credit distribution (Lukovits & Vinkler, 1995) | N | Y | N | Y | N | Y | N | Y | N | Y |
| Pure geometric counting (Egghe et al., 2000) | Y | N | N | N | N | N | N | N | N | N |
| Noblesse Oblige (Egghe et al., 2000; Zuckerman, 1968) | Y | N | N | N | N | N | Y | Y | N | Y |
| Refined weights (Trueba & Guerrero, 2004) | Y | Y | N | N | Y | N | N | Y | N | Y |
| Sequence determines credit (Tscharntke et al., 2007) | Y | N | N | N | N | N | NA | N | N | Y |
| Equal contribution (Tscharntke et al., 2007) | Y | N | N | N | N | N | NA | N | N | Y |
| Weight coefficients (Zhang, 2009) | Y | N | N | N | N | N | NA | N | N | Y |

| Counting method | Methods to support adequacy ||||| Elements that work against sensitivity || Elements that work against homogeneity |||
| --- | --- | --- | --- | --- | --- | --- | --- | --- | --- | --- |
| | Compare to other counting methods or bibliometric indicators | Principles to guide the definitions of the counting methods | Quantitative models for distributions of scores | Surveys or other empirical evidence | Compare groups of objects of study | Time specific evidence | Fixed credits for selected basic units of analysis | Parameter values selected by bibliometrician | External elements | Conditional equations |
| Golden productivity index (Assimakis & Adam, 2010) | Y | N | N | N | N | N | Y | N | N | Y |
| Tailor based allocations (Galam, 2011) | Y | N | N | N | N | N | N | Y | N | Y |
| Pareto weights (Tol, 2011) | Y | N | N | N | N | N | N | N | Y | N |
| Shapley value approach (Papapetrou, 2011) | Y | N | N | N | Y | N | N | N | Y | N |
| A-index (Stallings et al., 2013) | Y | Y | N | N | Y | N | N | N | N | N |
| Weighted fractional output, intra-mural or extra-mural (Abramo et al., 2013) | Y | N | N | N | Y | N | Y | N | N | Y |
| Absolute weighing factor (Aziz & Rozing, 2013) | Y | N | N | N | Y | N | N | N | N | Y |
| Contribution to an article's visibility, first approach (Egghe et al., 2013) | N | N | Y | N | Y | N | N | N | Y | N |
| Contribution to an article's visibility, second approach (Egghe et al., 2013) | N | N | Y | N | Y | N | NA | N | Y | N |
| [Shen & Barabasi, 2014] | N | N | N | N | Y | N | N | N | Y | N |
| Network-based model (Kim & Diesner, 2014) | Y | N | N | Y | Y | Y | N | Y | N | Y |
| Norwegian Publication Indicator (Sivertsen, 2016) | Y | N | Y | N | Y | N | NA | N | N | N |
| [Zou & Peterson, 2016] | Y | N | N | Y | Y | Y | NA | N | N | N |
| Relative intellectual contribution (Rahman et al., 2017) | Y | N | N | N | N | N | N | N | Y | N |
| Parallelization bonus (de Mesnard, 2017) | Y | Y | N | N | N | N | NA | N | N | N |
| [Steinbrüchel, 2019] | Y | N | N | N | N | N | N | N | Y | N |
| [Bihari & Tripathi, 2019] | Y | N | N | N | N | N | N | N | N | N |

### 3.3.1 Adequacy – five methods

According to the adequacy criterion, an indicator should be an adequate proxy for the object the indicator is designed to measure. Adequacy is tested through an independent and accepted measure for the object. The analysis identifies five methods in the studies that introduce the counting methods identified by RQ 1 that may be used to assess the adequacy of the counting methods:

- Compare to other counting methods or bibliometric indicators

- Principles to guide the definitions of the counting methods

- Quantitative models for distributions of scores

- Surveys or other empirical evidence

- Compare groups of objects of study

Below, the results report how well each of the methods assesses adequacy. There are examples of studies that explicitly use the methods to assess the adequacy of the counting methods. These include Sivertsen's use of a quantitative model (Sivertsen, 2016, p. 911) and Shen & Barabási's use of groups of objects of study comprised of Nobel laureates versus their co-authors (Shen & Barabasi, 2014, p. 12326). However, all the studies include analyses of the counting methods. These analyses include methods that, in the review, are interpreted as assessments of the internal validity of the counting methods.

**Compare to other counting methods or bibliometric indicators**

When the adequacy of counting methods is assessed by comparisons to other counting methods or bibliometric indicators, these other counting methods or bibliometric indicators should constitute independent and accepted measures of the aims of the counting methods. The aims of the counting methods are analyzed in relation to RQ 2 via Framework 2.

In the studies that introduce the counting methods, 25 of the 32 counting methods are analyzed by making comparisons with other counting methods or bibliometric indicators. An example are studies that introduce those counting methods for which the aim is that first- or last-authors provide the largest contribution / participation / … to a publication (Group 1 in Framework 2). Complete-fractionalized counting does not reflect this aim because all co-authors are credited equally. Therefore, comparisons involving the complete-fractionalized counting method that result in weak correlations may be regarded as assessments of the adequacy of the counting methods emphasizing first- or last-authors contributions (for examples, see Abramo et al., 2013, p. 207; Assimakis & Adam, 2010, pp. 424–425).

To successfully use comparisons with other counting methods or bibliometric indicators to assess adequacy, the bibliometrician should first evaluate the relevance of the other counting methods or bibliometric indicators in relation to aim of the counting method under assessment.

Furthermore, according to the adequacy criterion, not only should the other counting methods or bibliometric indicators used in comparisons be accepted measures of the aims of the counting methods, they should also be independent of the counting method under assessment. However, it can be debated whether or not other counting methods or bibliometric indicators are independent from the counting methods under assessment, as both build on publications and / or citations. Biases in the other counting methods or bibliometric indicators may also very well be present in the counting methods under assessment. This potential bias should be taken into account if adequacy of counting methods is assessed by comparisons to other counting methods or bibliometric indicators.

**Principles to guide the definitions of the counting methods**

When principles to guide the definitions of the counting methods are used to support the adequacy of the counting methods, these principles constitute an ideal description of the counting methods and, as such, represent independent and accepted measures of the aims of the counting methods. The aims of the counting methods are analyzed in relation to RQ 2 via Framework 2.

Six of the 32 counting methods have principles to guide their definitions. The six counting methods aim to measure (the impact of) the contribution / participation / … of an object of study (Group 1 in Framework 2). The principles are used to design the counting methods but not necessarily to assess the counting methods. Five out of the six studies emphasize the principles rank-dependency and / or fractionalization (Hodge & Greenberg, 1981; Lukovits & Vinkler, 1995, pp. 92–93; Stallings et al., 2013, pp. 9681–9682; Trueba & Guerrero, 2004, pp. 182–183). The remaining study's principles focus on division of tasks (de Mesnard, 2017).

Decisions about whether the principles to guide definitions of the counting methods are independent and accepted measures of the aims of the counting methods are in some cases based on thorough analyses (de Mesnard, 2017) and in other cases on personal experiences (Hodge & Greenberg, 1981). In the latter case, it is difficult to assess if the principles are appropriate for assessing the adequacy of the counting methods.

**Quantitative models for distributions of scores**

When quantitative models for distributions of scores are used to assess the adequacy of the counting methods, the scores for the objects of study are tested against distributions, which should constitute independent and accepted measures of the aims of the counting methods. The aims of the counting methods are analyzed in relation to RQ 2 via Framework 2.

Four of the 32 counting methods have quantitative models for distributions of scores. These distributions are: the Gini-coefficients close to 0.5 (Kyvik, 1989, pp. 209–210), Lotka's law (Egghe et al., 2013, pp. 59–62; Kyvik, 1989, p. 211), and equal scores across objects of study (Kyvik, 1989, pp. 207–208; Sivertsen, 2016, p. 911). For two of the counting methods, the aim is to give a balanced representation of productivity across research disciplines. This aim is reflected in the quantitative model (Kyvik, 1989, pp. 207–208; Sivertsen, 2016, p. 911). Whether the other quantitative models can work as independent and accepted measures of the aims of the counting methods depends on the validity of the models. Many studies in the bibliometric literature analyze Lotka's law, and the Gini-coefficient is investigated in economics, bibliometrics, and other research fields. As such, we have general knowledge about these quantitative models. This said, in the studies that use the models to assess the adequacy of counting methods, the relation between the aim of the counting methods and the model must be made clear.

**Surveys or other empirical evidence**

When surveys or other empirical evidence are used in the assessment of adequacy of counting methods, the credits to the basic units of analysis are evaluated against empirical data about how co-authors in a publication share credits. The idea is to define target values for the credits for the basic units of analysis. These target values should constitute independent and accepted measures of the aims of the counting methods. The aims of the counting methods are analyzed in relation to RQ 2 via Framework 2.

Four of the 32 counting methods use results from surveys (Kim & Diesner, 2014, pp. 593–595; Lukovits & Vinkler, 1995, pp. 93–94; Zou & Peterson, 2016, pp. 904–906), and one study uses a calculation of the ratio of senior researchers to non-senior researchers to determine credits for basic units of analysis (Boxenbaum et al., 1987, pp. 566–568). All four counting methods aim to measure (the impact of) the contribution / participation / … of an object of study (Group 1 in Framework 2).

Surveys and other empirical evidence can be well suited to create independent and accepted measures of the aims of counting methods. However, the study designs that inform the surveys and evidence are important to take into account. Certainly, surveys or other empirical evidence are created at a specific point in time, and this can impact the sensitivity of the counting methods (see Section 3.3.2).

**Compare groups of objects of study**

When comparisons between groups of objects of study are used to assess adequacy, scores for groups of objects of study are compared, for example early career versus senior researchers. These comparisons between groups of objects of study should rely on independent and accepted measures of the aims of the counting methods. The aims of the counting methods are analyzed in relation to RQ 2 via Framework 2.

Sixteen of the 32 counting methods use the compare groups of objects of study approach to assess the adequacy of the counting methods. The comparisons are between institutions or countries (for example, Howard et al., 1987), research fields or disciplines (for example, Sivertsen, 2016, p. 911), publication sets from different databases (for example, Shen & Barabasi, 2014, p. 12326), high-impact and other researchers (for example, Abramo et al., 2013, pp. 204–206), principal investigator and student (for example, Egghe et al., 2013, p. 64), or award-winners and other researchers (for example, Aziz & Rozing, 2013, pp. 4–6).

The studies that make comparisons between groups of objects of study to assess adequacy have different aims for introducing the counting methods. The studies represent all four groups from Framework 1. For some of the counting methods, the comparisons can be related to evaluations of the adequacy of the counting methods, for example, in comparisons of principal investigator versus student, where the expectation would be that the principal investigator would have the higher score (Egghe et al., 2013, p. 64). In other cases, the relation is less clear, and assessment of validity of adequacy may not be the intention of the studies making the comparisons.

### 3.3.2 Sensitivity – two elements

According to the sensitivity criterion, an indicator should reflect changes over time in the object the indicator is designed to measure. For counting methods, it is important that they are able to adapt to the increasing average number of co-authors per publication. The analysis identifies two elements in the counting methods identified by RQ 1 that make counting methods less adaptable to increasing numbers of authors per publication:

- Time specific evidence

- Fixed credits for selected basic units of analysis

Below, the results report for each of these two elements the effect resulting from an increasing number of authors per publication. The studies that introduce the counting methods do not analyze the issue of increasing numbers of authors per publication.

**Time specific evidence**

This element is overlapping with the method: surveys or other empirical evidence. As discussed in Section 3.3.1, counting methods where the adequacy is tested against the results of surveys or other empirical evidence about how co-authors of a publication share credits may eventually become obsolete due to changes in co-author practices not being reflected in the empirical evidence. One study creates new evidence (Zou & Peterson, 2016, pp. 904–906). However, some of the evidence from the four studies using surveys or other empirical data dates back to the 1980s (Boxenbaum et al., 1987, p. 567; Kim & Diesner, 2014, p. 594; Lukovits & Vinkler, 1995, p. 94; Vinkler, 1993, pp. 217–223). A further limitation of the evidence is that it relates to investigations of smaller numbers of authors per publication. For example, two studies include publications with up to five authors (Kim & Diesner, 2014, p. 594; Lukovits & Vinkler, 1995, p. 94). The four counting methods using empirical evidence aim to measure (the impact of) the contribution / participation / … of an object of study (Group 1 in Framework 2). To support this aim, the empirical evidence must be updated regularly to reflect the current average number of co-authors per publication.

**Fixed credits for selected basic units of analysis**

Three counting methods use fixed credits for selected authors only, independent of the number of co-authors. As the average number of co-authors increases, the credits for each of the other authors will decrease. The differences in credits assigned to the selected versus other authors may become extreme and, therefore, may not comply with the sensitivity criterion. The three counting methods that apply fixed credits for selected basic units of analysis (Abramo et al., 2013, p. 201; Assimakis & Adam, 2010, pp. 422–423; Egghe et al., 2000, p. 146) all aim to measure (the impact of) the contribution / participation / … of an object of study (Group 1 in Framework 2). The use of fixed credits may not reflect this aim if the average number of co-authors increases.

### 3.3.3 Homogeneity – three elements

According to the homogeneity criterion, an indicator should measure only one dimension and avoid heterogeneous indicators. The analysis investigates different elements that contribute to the equations for counting methods. Where there are several different elements in the equations, it is not immediately clear how these different elements affect the scores obtained by the counting methods. Such elements do not support homogeneity.

The analysis identifies three elements in the counting methods identified by RQ 1 that contribute to heterogeneity and, therefore, work against the homogeneity of the counting methods:

- Parameter values selected by bibliometrician

- External elements

- Conditional equations

Below are results for each of these three elements. None of the studies that introduce the counting methods analyze homogeneity.

**Parameter values selected by bibliometrician**

Seven of the 32 counting methods have one or more parameter values for the bibliometrician to select individually in each analysis. A change of parameter values will change the distribution of credits among the basic units of analysis of a publication. This ensures the counting methods can be adapted to accommodate credit distribution traditions in various research fields. For an example, see the illustration of 'Network-based model (Kim & Diesner, 2014)' in Figure 2 in Section 3.2.1.

The most common situation seen in the five counting methods is that the bibliometrician selects the value of a parameter, and that this value can vary between 0 and 1 (Egghe et al., 2000, p. 146; Ellwein et al., 1989, p. 321; Kim & Diesner, 2014, p. 591; Lukovits & Vinkler, 1995, pp. 92–95). Two counting methods include several parameter values to be selected by the bibliometrician (Galam, 2011, p. 371; Trueba & Guerrero, 2004, pp. 184–185). The effect of selecting a given parameter value as opposed to another parameter value is not immediately clear in the score obtained by the counting method; therefore, the counting method is heterogeneous.

**External elements**

The counting methods defined in Sections 3.2.1 and 3.2.2 as rank-dependent counting methods, in which the credits for basic units of analysis are shared unevenly based on characteristics other than rank, use external elements, for example, an author's position as principal investigator, an author's $h$-index, or an author's number of publications.

Seven of the 32 counting methods include external elements. In five counting methods, these external elements are author-level bibliometric indicators, for example the $h$-index. Sometimes the bibliometrician can chose between several indicators (Egghe et al., 2013, pp. 58–59; Papapetrou et al., 2011, pp. 554–555) and in other cases, specific indicators are used in the definitions of the counting methods (Shen & Barabasi, 2014, pp. 12325–12327; Tol, 2011, pp. 292–293). Other external elements are whether or not an author is a principal investigator (Boxenbaum et al., 1987; Steinbrüchel, 2019, pp. 307–308) and the type and extent of author contributions to a publication cf. the *CRediT* taxonomy[10] (Rahman et al., 2017, p. 278). At present, author contributions are not often an element made explicit by the publications included in an analysis.

---

[10] See Footnote 6.

However, it should be noted that, increasingly, journal publications include author contribution statements. External elements require background information about the author, and the effect of this background information on the score obtained by the counting methods is not immediately clear. Therefore, the counting methods are heterogeneous.

**Conditional equations**

Most counting methods have one equation, which is applied to all basic units of analysis and all publications. But some counting methods divide basic units of analysis or publications into groups according to specific characteristics and then use conditional equations on each group. Thirteen of the 32 counting methods apply conditional equations. Nine of the 13 counting methods use author rank, for example, first author, to divide the basic units of analysis into groups and apply conditional equations to the groups (for examples, see Assimakis & Adam, 2010, p. 422; Galam, 2011, p. 371; Trueba & Guerrero, 2004, pp. 184–185). Three counting methods use the number of authors per publication to create groups (Aziz & Rozing, 2013, p. 2; Kyvik, 1989, p. 206; Zhang, 2009, p. 416). In one counting method, one group has publications with first- and last authors from the same institution, and, in the other group, first- and last authors are from different institutions (Abramo et al., 2013, p. 201). These groupings of basic units of analysis or publications mean that it is not immediately clear how the counting methods' scores are obtained. Therefore, the counting methods are heterogeneous.

## 3.4   RQ 4: Three counting methods are used in four or more research evaluations

> RQ 4: To what extent are the counting methods identified by RQ 1 used in research evaluations and to what extent is this use compliant with the definitions in the studies that introduce the counting methods?

RQ 4 employs a literature search to identify research evaluations that use the counting methods identified by RQ 1. The focus is on research evaluations covering minimum 30 researchers. Some counting methods are used by the same author in several research evaluations. In such cases, only one of the research evaluations is counted. The Supplementary Material (see Section 7.1) provides a detailed schematic overview of the results related to RQ 4.

Fifteen of the counting methods are not used in research evaluations covering minimum 30 researchers and 14 counting methods are used in one to three research evaluations. Only three counting methods are used in four or more research evaluations: Harmonic counting, Hodge & Greenberg's counting method, and Sequence determines credit (Hodge & Greenberg, 1981; Howard et al., 1987; Tscharntke et al., 2007). For each of these three counting methods, a random sample of five research evaluations using the counting methods is selected for a further analysis of how the counting methods are used. Appendix 2 lists the research evaluations included in the analysis.

The research evaluations that use the three counting methods should draw on the same characteristics for the counting methods as presented at the introduction of the counting methods (see Section 3.2). If one or more of these characteristics change between the introduction and use of the counting method, then the research evaluation's use of the counting method may compromised.

In the samples of research evaluations using Harmonic counting and Sequence determines credit, the counting methods are used with the same characteristics as in the studies that introduce the counting methods. As in the introduction of the counting methods, the objects of study and the basic units of analysis are authors and the arguments for the use of the counting methods are from Group 1: 'The indicator measures the (impact of) contribution / participation / … of an object of study'.

The research evaluations' arguments for using Howard et al.'s counting method are also from Group 1, a situation which is in agreement with the study introducing the counting method. However, in the introduction of the counting method, authors are the basic units of analysis and institutions are the objects of study. In the research evaluations, the basic units of analysis and the objects of study are authors, institutions, or countries. The research evaluations that have basic units of analysis and objects of study other than those present at the introduction of the counting method do validate the counting methods by comparisons with other counting methods (see Section 3.3.1 for more about this method for assessment of adequacy). However, one study that has countries as objects of study does not validate the counting method at all (Tsai & Lydia Wen, 2005). In this research evaluation, the use of Howard et al.'s counting method may be compromised as the use is not validated by either the study introducing the counting method or by the research evaluation using the counting method.

## 3.5 Summary of results

Section 3 presents the results related to RQs 1 – 4. The results can be summarized as follows. The review identified 32 counting methods introduced during the period 1981 – 2018. No unique counting methods are introduced during the earlier period 1970 – 1980. Two frameworks categorize these counting methods. Framework 1 describes selected mathematical properties of counting methods, and Framework 2 describes arguments for choosing a counting method. Twenty of the 32 counting methods are rank-dependent, fractionalized, and introduced to measure the (impact of) the contribution / participation / … of an object of study (Group 1 in Framework 2). Next, three validity criteria are used to identify five methods that test the adequacy of counting methods, two elements that test sensitivity, and three elements that test homogeneity of the counting methods. These methods and elements are from the studies that introduce the counting methods and may be used to assess the internal validity of counting methods. Finally, a literature search finds research evaluations that use the counting methods. Only three of the 32 counting methods are used by four research evaluations or more. Of these three counting methods, two are used as described in the studies that introduce the counting methods.

# 4 Discussion

Section 4 has two parts. Section 4.1 discusses the methods used in the present review and the limitations resulting from the methodological choices made. Section 4.2 gives interpretations of the results presented in the review.

## 4.1 Discussion of the methods - Limitations

Section 4.1 discusses the review's methods and their limitations. Results related to an RQ inform subsequent RQs (see Table 1, Section 2); therefore, limitations related to the methods used in relation to RQ 1 have consequences for all the following RQs, and the limitations for RQ 2 affect RQ 3 and 4.

### 4.1.1 RQ 1: Literature search covers counting methods in the bibliometric research literature

The literature search aimed at identifying counting methods and undertaken in relation to RQ 1 forms the basis for the review. The literature search covers peer-reviewed studies in English from the period 1970 – 2018. Including more publication types and more languages could lead to the identification of additional counting methods, for example, counting methods used in local university reports. To include the period 2019 – 2020 would most likely result in more counting methods; however, counting methods introduced after 2018 are excluded from the review because the use of these will be difficult to assess in relation to RQ 4, i.e. less than two years after their introduction into the bibliometric research literature.

The literature search identifies 32 counting methods, which, in relation to RQ 2, are then assigned to categories. Including the period 2019 – 2020 could add counting methods to the analysis and; thus, more detail to the results. However, the proportion of counting methods in each of the categories is consistent over time, in that none of the categories include counting methods from one decade alone (see the Supplementary Material, Section 7.1). This suggests that adding a few new counting methods would be unlikely to change the overall results.

Even though the result of the literature search may be supplemented with more counting methods, the review scrutinizes more counting methods than previous reviews. The present review demonstrates that the majority of the counting methods are introduced in the most recent decade. If this trend continues, future counting methods may change the proportion of counting methods in each of the categories used to structure this review.

### 4.1.2 RQ 2: Two frameworks selected among other possible frameworks

RQ 2 uses two selected frameworks to categorize the counting methods identified by RQ 1 according to the characteristics of those counting methods. Framework 1 describes selected mathematical properties of counting methods and Framework 2 describes arguments for choosing a counting method for a bibliometric analysis.

The literature search did uncover other frameworks that may be suitable for the analysis of many different counting methods. Indeed, drawing on theories, methods, and concepts from other research fields, the number of potentially relevant frameworks is very large. However, using only one or a few frameworks in an analysis serves to prevent overly complex results. The present review uses two frameworks. For illustration is the potential of a framework not used in the review is discussed below.

Xu et al.'s framework divides counting methods into linear, curve, and "other" counting methods (Xu et al., 2016, pp. 1974–1977). A closer look at the counting methods in the present review with regard to what Xu et al. define as curved counting methods reveals that Zou & Peterson's counting method (Zou & Peterson, 2016, p. 906) is the non-fractionalized version of pure geometric counting (Egghe et al., 2000, p. 146). In both counting methods, the second author gets half the credit given to the first author, the third author gets half the credit given to the second author, and so on. Furthermore, Howard et al.'s counting method (Howard et al., 1987, p. 976) has similar characteristics. The credits are reduced by one third going from the first author to the second author, and so on. A further analysis utilizing Xu et al.'s framework may reveal other similarities that do not emerge from applying Frameworks 1 and 2. However, Xu et al.'s framework does not fit with as many counting methods as Frameworks 1 and 2 as 'citation-based credit assignment methods' — for example, Pareto weights — are not included in Xu et al.'s framework (Xu et al., 2016, p. 1974).

The two frameworks selected for the review represent a further development of the well-known dichotomy full versus fractional counting (Framework 1) and focus on the argument for the introduction of the counting methods (Framework 2). Thus, the frameworks illustrate that different approaches can be used to describe counting methods. Framework 1, Xu et al.'s framework, and other frameworks used in relation to counting methods draw on mathematical properties, which are highly relevant for the analysis of counting methods. However, applying Framework 2 in the review shows that approaches other than mathematical can add useful nuance to our knowledge about counting methods.

### 4.1.3   RQ 3: Homogeneity criterion may be developed further

RQ 3 applies three validity criteria (adequacy, sensitivity, and homogeneity) that are developed for and tested on bibliometric indicators (Gingras, 2014, pp. 116–119; Wildgaard, 2015a, sec. 6.3, 2019, sec. 14.4.1). While guidance for how to apply the criteria exists at the overall level implementation in a specific case, such as the present review, requires several choices.

The homogeneity criterion is difficult to use in relation to counting methods. The criterion guidance explains that a mix of different elements with the same measure unit does not present as heterogeneous but as composite (Gingras, 2014, p. 122). However, the difference between heterogeneous and composite is described with a less than ideal example (ibid.) that makes accurate interpretation of the guidance difficult.

The homogeneity criterion may be interpreted more strictly than is done in the review (see example in Section 2.3), leading to the definition of fewer elements for indicating heterogeneous counting methods. Or the difference between heterogeneous and composite may be ignored, leading to the definition of more elements for indicating heterogeneous counting methods.

### 4.1.4   RQ 4: Selective focus on peer-reviewed research evaluations

RQ 4 conducts a literature search to identify research evaluations that use the counting methods identified by RQ 1. This means that research evaluations that do not cite studies identified by RQ 1 are not found in the RQ 4 literature search. For some of the well-known counting methods, this could result in underrepresentation in the RQ 4 search results: research evaluations may mention the name of the counting method without a reference at all, or they may cite later studies describing the counting method rather than the original study that introduced the counting method.

Three counting methods identified by RQ 1 have several names and / or studies that introduce the counting methods. This is sometimes seen in bibliometric studies and can lead to misinterpretations (Gauffriau et al., 2008, pp. 166–169). The three counting methods are: Harmonic counting, Proportional counting (also known as Arithmetic counting), and Noblesse Oblige (Egghe et al., 2000, p. 146; Hagen, 2008; Hodge & Greenberg, 1981; Van Hooydonk, 1997; Zuckerman, 1968). A literature search of the alternative names and / or studies that introduce the counting methods is conducted. The search leads to a few additional research evaluations that use the counting methods identified by RQ 1. In the results related to RQ 4, Proportional counting would change from the interval 'zero research evaluations use the counting method' to 'one to three research evaluations use the counting method'. This change would have no impact on the results of the review (see the Supplementary Material, Section 7.1).

Furthermore, the review excludes research evaluations in reports and gray literature and only includes peer-reviewed studies in English. These limitations mean that results regarding the use of the counting methods may be underreported. A broader literature search could be conducted for selected languages or by introducing limitations other than the ones chosen in the present review to manage the literature search. It is worth noting, however, that including all publication types and all languages would be impractical, resulting in a huge search. The result set of such a search would pose considerable challenges for analysis, especially for qualitative approaches such as those employed in relation to RQ 4.

A final point is that the focus of the review is on the applied use of counting methods in larger research evaluations, and thereby, the scalability of the counting methods. This focus can be adjusted to accommodate other types of use cases, for example, investigations of how existing counting methods inform the development of new counting methods or tests of the mathematical properties of the counting methods. As with the choice of frameworks for categorizations of counting methods, there are many possibilities for analyses exploring the use of counting methods.

## 4.2  Discussion of the results

Section 4.1 discussed the review's methods and their limitations. Section 4.2 discusses interpretations of the results related to RQs 1 – 4.

### 4.2.1  RQ 1: New introductions of counting methods underline the relevance of analyses of counting methods

The results related to RQ 1 show that counting methods in the bibliometric research literature should neither be reduced to a simplified choice between two counting methods — full and fractional counting — nor be implicit in bibliometric analyses. The literature search identifies 32 counting methods, and the majority (17 counting methods) is introduced in the most recent decade, a situation that underlines the relevance of the review.

### 4.2.2  RQ 2: Consistent analyses of counting methods reveal categories of counting methods

The results related to RQ 2 demonstrate that consistent analyses of counting methods provides new knowledge and allows a more nuanced understanding of counting methods.

Below, three main observations based on the results of applying Frameworks 1 and 2 are discussed. The observations do not relate to counting methods introduced in a specific decade; rather they are valid for counting methods from the 1980s as well as for counting methods from the 2010s. Following the three observations, counting methods not fitting the frameworks are discussed.

**Observation 1**

The first observation is that all counting methods are introduced with specific basic units of analysis and objects of study, often authors. Recall that complete counting and whole counting methods are identical at the micro-level but often result in different scores at the meso- or macro-level. In other words, this difference is not visible if two counting methods are only defined at the micro-level. Thus, not all definitions of the counting methods necessarily hold if the counting methods are applied at aggregation levels other than the aggregation levels for which they are specifically defined (often the micro-level). The use of the counting methods would be facilitated if they were to be introduced as score functions, which could be combined with different basic units of analysis and objects of study.

**Observation 2**

The second observation is about rank-dependent counting methods, excluding rank-dependent counting methods based on other characteristics than rank. The majority (19 of the 32 counting methods) are rank-dependent. Again, most of these counting methods are defined at the micro-level.

Among the pre-1970 counting methods, straight counting is rank-dependent. Older studies have shown that the difference in scores obtained by straight counting and other pre-1970 counting methods levels out at the meso- and macro-level. Obviously, when straight counting is used at the micro-level, it is important to be the first author of a publication in order to receive credit for that publication (Lindsey, 1980, pp. 146–150). However, at the meso- or macro-level, straight counting scores for institutions or countries are fair approximations of scores resulting from whole or fractional counting (Braun et al., 1989, p. 168; Cole & Cole, 1973, pp. 32–33).

As discussed in Section 1.1, today, there are more institutions and countries per publication. Therefore, in analyses with recent publications, scores resulting from whole versus straight counting are more likely to differ (for examples, see Gauffriau et al., 2008, pp. 156–157; Lin et al., 2013). However, straight and complete-fractionalized counting still yield similar scores at the meso- or macro-level (ibid.).

Given the situation described above, it is likely that the results regarding straight counting would hold for other rank-dependent and fractionalized counting methods. In theory, this means that the 14 rank-dependent and fractionalized counting methods could be substituted by complete-fractionalized counting for analyses at the meso- and macro-level. Complete-fractionalized counting is rank-independent and, therefore, easier to apply compared to rank-dependent counting methods with their more complex equations. In future research, it would be interesting to investigate comparisons between complete-fractionalized counting and the 14 rank-dependent and fractionalized counting methods using empirical data at the meso- and macro-level.

**Observation 3**

The third observation is that almost all of the counting methods (28 of 32) are introduced with an argument from Group 1 in Framework 2: 'The indicator measures the (impact of) contribution / participation / … of an object of study'. This result suggests that the common understanding of counting methods is that they relate to the concept that the study aims to measure, for example, the objects of study's participation in a research endeavor as measured by whole counting. However, four of the counting methods in the review show that there are alternative arguments for introducing counting methods (see the Supplementary Material, Section 7.1). An interpretation that assumes all counting methods aim to measure the contribution / participation / … of an object of study would be a mistake.

**Counting methods that do not fit the study frameworks**

In addition to the observations above, as shown in Section 3.2.5, two counting methods do not fit the selected properties from Framework 1. Also, as Section 3.2.6 illustrates, two arguments for introducing counting methods do not currently comply with Framework 2. In the context of the present review, it is no surprise that not all counting methods fit Frameworks 1 and 2. As discussed in Section 4.1.2, neither does Xu et al.'s framework cover all counting methods.

Indeed, the counting methods outside the frameworks offer a potential opportunity to investigate the further development of the frameworks. As well, an analysis of why the counting methods do not fit the frameworks could give new perspectives on the counting methods. To this end, Appendix 1 presents an example case study using Frameworks 1 and 2 to analyze the counting method used in the Norwegian Publication Indicator (NPI), and through this analysis, to identify potential for developing the frameworks further and achieving deeper understanding of the NPI.

### 4.2.3 RQ 3: Assessment of internal validity of counting methods can be developed further

The results related to RQ 3 present the application of three criteria to evaluate the internal validity of counting methods. Firstly, methods to assess the adequacy of counting methods, and thus, to support the internal validity of the counting methods are presented. Next, the analysis considers elements that define weak sensitivity of the counting methods. Finally, elements in counting methods that make those counting methods heterogeneous are examined. Elements connected with weak sensitivity and heterogeneous elements do not support the internal validity of counting methods.

The use of the adequacy criterion in relation to counting methods suggests that adequacy can be analyzed in relation to the aims of the counting methods, i.e. the arguments for introducing the counting methods from Framework 2. The use of the sensitivity criterion on counting methods shows that seven counting methods have elements that indicate weak sensitivity (see the Supplementary Material, Section 7.1). On the other hand, the remaining counting methods do not accommodate explicit measures to support sensitivity, i.e. reflecting the increasing number of authors per publication over time. Likewise, the use of the homogeneity criterion in relation to counting methods indicates that a large number (21 of the counting methods or 15 counting methods if a strict interpretation of the criterion is used, see Section 4.1.3) have heterogeneous elements that work against homogeneity. At least for these counting methods, no specific measures are taken to support homogeneity. Thus, the results related to RQ 4 suggest potential for the consistent use of validity criteria in relation to counting methods.

The bibliometrician can decide not to use counting methods with elements that work against sensitivity and homogeneity (see the Supplementary Material, Section 7.1). However, there may be more elements than those identified in the review, potentially leading to the exclusion of more counting methods. Alternatively, the heterogeneous counting methods may have high adequacy and thus prove useful in research evaluations anyway. As such, the present review's application of the three criteria for validity should be used as attention and guiding points for selecting counting methods — but not as a selection key. Wildgaard reaches a similar conclusion after her implementation of the three criteria (Wildgaard, 2015a, p. 95).

### 4.2.4 RQ 4: The context in which the counting method are used should be assessed

The results related to RQ 4 investigate to what extent the counting methods identified by RQ 1 are used in research evaluations, and whether research evaluations use the counting methods in agreement with how the counting methods are described initially in the studies that introduce them. The analysis finds that a large majority of the counting methods (29 of 32) are either used in maximum three research evaluations or not used at all in research evaluations. The paradox of this moderate use and new counting methods continuously being introduced into the bibliometric research literature remains unsolved.

Three counting methods are used in at least four research evaluations. In one instance, a counting method is used in a research evaluation with other basic units of analysis and objects of study than those defined in the introduction of the counting method. It is important to be aware of the contexts in which the counting methods are used and whether these contexts differ from the definition of the counting methods. In a previous study, the results show that the use of pre-1970 score functions are not consistent across studies. Many of the score function are used with several arguments from Framework 2 in the bibliometric research literature (Gauffriau, 2017).

### 4.3 Summary of discussion

Section 4 discusses the review's methods and, thus, limitations of the results. The section continues with interpretations of the review's results.

The review uses a range of methods, starting with a literature search to identify counting methods, followed by the application of frameworks for categorizations of the counting methods, the implementation of validity criteria for the counting methods, and finally, an investigation exploring to what extent the counting methods are used in research evaluations and whether the research evaluations use the counting methods in agreement with how the counting methods are described in the studies that introduce them. The selected frameworks for categorizations of the counting methods particularly impact the results, as other frameworks could have been applied, thereby changing the focus of the results. As well, the review's focus on the use of counting methods in research evaluations could be altered, for example, to focus instead on other types of use cases. It has not been possible to cover all frameworks for categorizations of the counting methods and all use cases in the same review. However, the review may inspire new studies of counting methods using other frameworks, methods, use cases, etc.

The results presented in the review lead to several observations that point to new knowledge about counting methods. Most of the counting methods are introduced in the most recent decade, which highlights the timeliness of the review. It seems that the common understanding of counting methods is that they relate to the concept that the study using them attempts to measure (Group 1 in Framework 2). Most of the counting methods are defined at the micro-level and not as score functions. Score functions can be combined with different basic units of analysis and objects of study, and this in turn can facilitate use. Also, most of the counting methods are rank-dependent and, therefore, they may only be relevant for use at the micro-level. Finally, only three of the counting methods are used in research evaluations. This presents a paradox given that new counting methods are continuously introduced into the bibliometric research literature.

# 5   Conclusion

This review shows that the topic of counting methods in bibliometrics is complex. However, in bibliometric analyses, counting methods are often implicit or reduced to a binary choice between full and fractional counting. Furthermore, bibliometric handbooks and textbooks do not discuss counting methods, cover only a fraction of the available counting methods, and / or lack well-defined frameworks for describing and categorizing the counting methods.

To design an indicator, a bibliometrician must chose a counting method. In many cases, this choice will affect the score obtained by an object of study. Thus, the study of counting methods is essential to the understanding of bibliometric indicators and to the informed choice of these indicators in practice.

The review shows that there are at least 32 counting methods introduced into in the bibliometric research literature after 1970. Other reviews of counting methods do not cover as many counting methods. The counting methods can be categorized according to their characteristics.

The categorizations of counting methods in the review lead to several new findings. One important finding is that 28 of the 32 counting methods covered by the review are defined at the micro-level (authors). This makes it difficult to use these counting methods at other aggregation levels in a consistent manner. Another important finding suggests that the common understanding of counting methods is that they relate to the concept that the study using the counting methods aims to measure. Furthermore, the review applies three internal validity criteria for well-constructed bibliometric indicators (adequacy, sensitivity, and homogeneity) to counting methods for the first time. The criteria help identify methods and elements useful for assessing the internal validity of counting methods. Finally, the review documents the paradox between the many counting methods introduced into the bibliometric research literature and the finding that only few of these counting methods are used in research evaluations.

The review provides practitioners in research evaluation and researchers in bibliometrics with a detailed foundation for working with counting methods. At the same time, many of the findings in the review provide bases for future investigations of counting methods. Well-defined frameworks other than those used in the review could be applied to investigate counting methods. The categories of counting methods identified in the review could also be analyzed further, for example, through a study of how to use the counting methods at different aggregation levels. A further evaluation could be carried out of the methods and elements deemed useful for assessing internal validity of counting methods. And, finally, the use of counting methods in contexts other than research evaluations could be examined. The schematic overview of the results of the review presented in the Supplementary Material (see Section 7.1) may be a useful starting point for inspiring further investigations of counting methods.

# 6   Acknowledgements

The author would like to thank Dr. Lorna Wildgaard for critically reading earlier versions of the manuscript, Dr. Dorte Drongstrup for fruitful discussions in relation to the manuscript, and Dr. Abigail McBirnie for proof-editing.

# 7 Supplementary Material and Appendices

Section 7 includes:

- Introduction to the Supplementary Material, which offers a schematic overview of the results related to all RQs.

- Appendix 1, which presents a case study of the Norwegian Publication Indicator (NPI).

- Appendix 2, which provides references to the research evaluations included in the analysis related to RQ 4.

## 7.1 Introduction to the Supplementary Material

The Supplementary Material offers a schematic overview of the results under all RQs saved in an excel file: https://static-curis.ku.dk/portal/files/252611821/Suppl_Material_Counting_methods_1970_2018_review.xlsx

The excel file allows the reader to easily filter or sort the four pre-1970 counting methods as well as the 32 counting methods identified by RQ 1. Thus, the excel file invites the reader to explore the counting methods presented in this review, their characteristics, methods and elements to assess their internal validity, and their use in research evaluations. Such exploration may inspire new studies of counting methods, leading to even more new knowledge about counting methods.

## 7.2 Appendix 1: Case study of the Norwegian Publication Indicator (NPI)

The results related to RQ 2 report that the Norwegian Publication Indicator (NPI) does not fit with either of the two frameworks used to categorize the counting methods identified by RQ 1. The NPI (Sivertsen, 2016) was published after Framework 1 (Gauffriau et al., 2007) and was not included in the sample of studies analyzed for the development of Framework 2 (Gauffriau, 2017). Appendix 1 presents a case study that analyze the conflicts between the NPI and the two frameworks. The aim of the analysis is to provide more insight into the NPI and, potentially, to develop the frameworks further.

### 7.2.1 The NPI counting method

The NPI is used in Norway's national performance-based funding model. On the basis of this model, 2% of the basic funding for the higher education sector is distributed according to the NPI (Sivertsen, 2018, p. 9). The indicator's calculation involves five steps (Norwegian Center for Research Data et al., 2018):

1. Calculate the total number of author shares in the publication. An author's share is any unique combination of authors and institutions in the publication.

2. Calculate how many author shares the institution has, and divide by the total number of author shares.

3. Calculate the square root of the number (fraction) from point 2.

4. Multiply by points for the level and type.

5. Multiply by 1.3 if the publication has affiliations to foreign institutions.

The NPI calculation above was implemented after an evaluation of the indicator in 2014. Balanced scores across disciplines were mentioned in the mandate for the evaluation and the evaluation report concluded (Aagaard et al., 2014, p. 8): "... the relatively large differences in average publication points per researcher appear to imply an unintentional redistribution of funding across disciplines. While there may be a number of factors behind these differences, the most important factor appears to be fractionalization rules based on the number of authors." This conclusion was supported by results in the report (Aagaard et al., 2014, pp. 57–62) and by a study published prior to the evaluation (Piro et al., 2013).

To create balanced scores across disciplines, step 3 of the NPI calculation was introduced: "We found that the balance is reached by applying the square root of the institutional fraction of the publication" (Sivertsen, 2016, p. 912). Other studies and a memorandum from the National Board of Scholarly Publishing reach similar results regarding the new counting method (Det nasjonale publiseringsutvalget, 2015; Sivertsen, 2018, pp. 11–12). Furthermore, Sivertsen et al. build on the NPI to investigate a counting method where not only the square root but also the cubic root, fourth root, etc. are applied as sensitivity parameters. The study evaluates the counting method using empirical data and concludes that the square root and cubic root preform well as sensitivity parameters (Sivertsen et al., 2019).

Steps 1 – 3 of the NPI are of particular interest to the present review as neither applying the square root to scores obtained by complete-fractionalized counting nor arguing for balanced scores across disciplines are covered in Frameworks 1 and 2. As such, the focus is on steps 1 – 3 of the NPI calculation in relation to assessing whether or not the frameworks used in the review can be used to analyze the NPI. Changes to steps 4 and 5 of the NPI calculation do not solve the issue of balanced scores across disciplines (Det nasjonale publiseringsutvalget, 2015, p. 2). Thus, the analysis does not include these two steps.

### 7.2.2 Analysis of the NPI using Framework 1

The two first steps in the calculation of the NPI are identical to complete-fractionalized counting with author shares (unique combinations of authors and institutions in a publication) as the basic units of analysis and institutions as objects of study. Complete-fractionalized counting complies with Framework 1. The third step of the NPI, however, is not covered by the framework. Applying the square root as done in the NPI does not comply with measure theory (von Ins, personal communication, 30[th] October, 2018), which is the theoretical basis for Framework 1.

To illustrate the conflict, a publication with three authors serves as an example. Author A and author B are from institution Y, and author C is from institution Z. According to the NPI, the two institutions' scores are $Y_1 = \sqrt{(1/3+1/3)} \approx 0.82$ and $Z_1 = \sqrt{(1/3)} \approx 0.58$. To comply with measure theory, the basic units of analysis must be credited individually, i.e. as mentioned in Section 2.2.1. In the NPI, this is not done for the basic units of analysis (author shares) from institution Y in the example above. To credit the basic units of analysis individually, the following change in the calculation can be applied: $Y_2 = \sqrt{(1/3)} + \sqrt{(1/3)} \approx 1.15$ and for $Z_2 = \sqrt{(1/3)} \approx 0.58$. However, this change will also change the scores obtained by the counting method.

The calculation of the NPI means that the properties related to Framework 1 do not necessarily hold for the NPI. In the framework, objects of study are scored by collecting credits from the basic units of analysis. The basic units of analysis are credited individually. The NPI does not meet this condition.

However, an empirical study of 14,441 Norwegian researchers' publications from the period 2011 – 2012 evaluates the adequacy of the NPI. The counting method proves successful in producing scores that satisfied balance across disciplines (Sivertsen, 2018, pp. 11–12, 2016, pp. 911–912). Thus, the use of the square root in the NPI may offer new knowledge about counting methods.

More research is needed to understand the theoretical foundation of the NPI counting method. The remaining part of Section 7.2.2 offers an analysis to better understand the NPI. Other approaches may add more insight into the theoretical foundation of the NPI.[11]

**The square root in the NPI**

There are bibliometric studies presenting uses of the square root that are similar to the use of the square root in the NPI. Examples include a study of the *Academic Ranking of World Universities* (ARWU) (Shanghai Ranking Consultancy, 2018) and a blog post about the *World University Rankings* (WUR) (Times Higher Education, 2018). Both of these university rankings use square root transformations. These two examples illustrate a possible theoretical foundation for the counting method used in the NPI.

In the ARWU, the rationale for using a square root transformation is: "the distribution of data for each indicator is examined for any significant distorting effect and standard statistical techniques are used to adjust the indicator if necessary" (according to Docampo, 2013 who is citing; N. C. Liu & Cheng, 2005). Liu & Cheng do not elaborate how this is done, but Docampo shows that "once the highest scoring institution is identified, [i.e.] the one with the largest number of HiCi [highly cited] authors, relative scores of the other institutions are calculated not in direct proportionality of the number of HiCi authors, but in direct proportionality between the square roots of those numbers" (Docampo, 2013, p. 569; square brackets added).

The WUR draws on *Web of Science* data, and prior to 2015, a 'regional modification' was used according to a blog post (Baty, 2015). It is not stated how the modification was made; however, a 2014 guideline to *InCites Indicators* describes the 'Normalized citation impact - country adjusted' indicator, and, in relation to that indicator, mentions the square root. A blog post about the WUR and the *InCites Indicators* guideline are quoted below to show the overlap between the two texts. Taken in combination, the two texts indicate that a square root transformation was used in the WUR before 2015. First, the blog post about WUR explains (Baty, 2015; quotation marks and italics from the original text are kept):

"Thomson Reuters explained: "*The concept of the regional modification is to overcome the differences between publication and citation behaviour between different countries and regions. For example some regions will have English as their primary language and all the publications will be in English, this* [sic] *will give them an advantage over a region that publishes some of its papers in other languages (because non-English publications will have a limited audience of readers and therefore a limited ability to be cited). There*

---

[11] Sivertsen et al. include an approach with a mathematical analysis of how the credit for an author is affected by a change in the rank of that author in the affiliation section of a publication or a change in the number of co-authors (Sivertsen et al., 2019, pp. 691–693). The present review does not cover this approach; however, the approach would be interesting to apply to the counting methods covered by the present review.

*are also factors to consider such as the size of the research network in that region, the ability of its researchers and academics to network at conferences and the local research, evaluation and funding policies that may influence publishing practice.""*

Second, the *InCites Indicators* guideline states (Thomson Reuters, 2014, p. 21; square brackets added):

"This [regional modification] is a modification of the normalized citation impact to take into account the country/region where the institution is based. This reflects the fact that in [*sic*] some regions will have different publication and citation behaviour because of factors such as policy, language and size of the research network. The indicator is calculated as the normalized citation impact of the institution divided by the square root of the normalised [*sic*] citation impact of the country in which it is based."

The use of square root in the NPI can be interpreted as comparable to the examples quoted above, and thus, as a data transformation method. This means that, in relation to Framework 1, the NPI can be interpreted as using complete-fractionalized counting and as applying the square root to the scores to transform those scores.

**Data transformations**

As discussed in Osborne's introduction on how to use data transformations (Osborne, 2012), a data transformation should be designed with the interpretation of the transformed data in mind. The order of the data values is kept, but the distances between values may change (Osborne, 2012, pp. 3–4). The argument can be made that this is the desired effect of applying the square root transformation in the NPI as doing so resolves the issue of balanced scores across disciplines. The importance of selecting a minimum value of a distribution is also highlighted, as it can be shown that the minimum value influences the efficacy of a transformation (Osborne, 2012, pp. 4–5). In the case of square root transformations for numbers between 0 and 1, as is the case in the NPI, the square root will increase the numbers, while for numbers above 1, the square root will decrease the numbers (Osborne, 2012, p. 3). Transformations for numbers between 0 and 1 are mentioned by Sivertsen (Sivertsen, 2016, p. 912). However, it is beyond the scope of this case study to analyze the NPI data and discuss other data transformations.

Bibliometric studies also use the square root for purposes other than data transformations. Selected examples include: distributions of bibliometric data (W. Glänzel & Schubert, 1985); selection of highly cited publications (Vinkler, 2009); random error estimation (Shearer & Moravscik, 1979); and, variations of the *h*-index that use the square root, for example, to calculate the geometric mean in the *hg*-index (for more examples, see Wildgaard, 2015b; Wildgaard et al., 2014). It is beyond the scope of this case study to investigate these other uses of the square root.

### 7.2.3   Analysis of the NPI using Framework 2

In Sivertsen's study about the change of the counting method in the NPI, the reason given for the change is that the NPI "needs to be balanced because it is used for measuring productivity across institutions with different research profiles (e.g. general versus technical universities, universities with and without medical faculties)" (Sivertsen, 2016, p. 911). The argument for the NPI is that it should give a balanced representation of productivity across research disciplines (Sivertsen, 2016, p. 912). As noted in Section 3.2.3, Kyvik makes a similar argument for the introduction of his counting method (Kyvik, 1989, p. 207).

Framework 2 does not cover the argument for the NPI. The incompleteness of the framework was anticipated, i.e. because the introduction of new indicators may well introduce new arguments for counting methods (Gauffriau, 2017, p. 678). The framework can beneficially be developed to include the argument for the NPI. Group 2 in the framework is 'Additivity of counting method', which is also the only argument in the group. The group can be expanded to include "Balanced scores across disciplines" and renamed "Mathematical properties of counting methods" as highlighted in Table 5.

Table 5: Updated categorization of arguments for counting methods for publication and citation indicators.

| Category | Counting method(s) |
|---|---|
| **Group 1: The indicator measures the (impact of)…** | |
| … participation of an object of study | Whole |
| … production of an object of study | Whole, complete-fractionalized |
| … contribution of an object of study | Whole, complete-fractionalized (rank-independent and rank-dependent) |
| … output / volume / creditable to / performance of an object of study | Whole, complete-fractionalized |
| … the role of authors affiliated with an object of study | Straight, last author, reprint author |
| **Group 2: ~~Additivity of counting method~~ Mathematical properties of counting methods** | |
| Additivity of counting method | Whole, complete-fractionalized |
| Balanced scores across disciplines | Norwegian Publication Indicator, steps 1–3 |
| **Group 3: Pragmatic reasons** | |
| Availability of data | Whole, straight, reprint author |
| Prevalence of counting method | Whole |
| Simplification of indicator | Whole |
| Insensitive to change of counting method | Whole |
| **Group 4: Influence on / from the research community** | |
| Incentive against collaboration | Complete-fractionalized |
| Comply with researchers' perceptions of how their publications and/or citations are counted | Whole |

### 7.2.4 Discussion and conclusion of the NPI case study

Using Frameworks 1 and 2 to analyze the NPI suggests that this counting method is a new type of counting method.

The NPI does not comply with measure theory. Thus, Framework 1 and the mathematical properties described in the framework do not apply to the NPI. An alternative interpretation is that the NPI is using complete-fractionalized counting (included in Framework 1) and is applying the square root to the scores to perform a data transformation of the scores.

Even though Framework 1 does not fit with the NPI, the framework still proves useful in the analysis of the NPI. The analysis shows how the NPI differs from other the counting methods identified by RQ 1. Interestingly, the NPI is defined with authors (or author shares, to be precise) as basic units of analysis and institutions as objects of study. If authors had been basic units of analysis and objects of study — which is the case for the introduction of most counting methods — then the special properties of the NPI would not have been visible. However, it remains unclear how to apply the NPI to other objects of study, for example, countries.

Neither is the NPI covered by Framework 2; however, this framework can easily be developed to include the argument for the NPI. Group 2 in Framework 2 can be broadened to include additional mathematical properties, i.e. balanced scores across disciplines, alongside additivity.

### 7.3 Appendix 2: Research evaluations analyzed under RQ 4

Appendix 2 lists the studies included in the analysis undertaken in relation to RQ 4. The analysis investigates to what extent the use of counting methods is compliant with the definitions of the counting methods in the studies that originally introduce the counting methods.

The random sample of five research evaluations that use Harmonic counting (Hodge & Greenberg, 1981):

- Jung, S., & Yoon, W. C. (2019). Citation-Based Author Contribution Measure for Byline-Independency. *2019 IEEE International Conference on Big Data (Big Data)*, 6086–6088. https://doi.org/10.1109/BigData47090.2019.9006230

- Lee, Jongwook, & Yang, Kiduk. (2015). Co-authorship Credit Allocation Methods in the Assessment of Citation Impact of Chemistry Faculty. 한국문헌정보학회지, *49*(3), 273–289. https://doi.org/10.4275/KSLIS.2015.49.3.273

- George, S., Lathabai, H. H., Prabhakaran, T., & Changat, M. (2020). A framework towards bias-free contextual productivity assessment. *Scientometrics*, *122*(1), 127–157. https://doi.org/10.1007/s11192-019-03286-7

- Wang, F., Jia, C., Wang, X., Liu, J., Xu, S., Liu, Y., & Yang, C. (2019). Exploring all-author tripartite citation networks: A case study of gene editing. *Journal of Informetrics*, *13*(3), 856–873. https://doi.org/10.1016/j.joi.2019.08.002

- Hagen, N. T. (2014). Counting and comparing publication output with and without equalizing and inflationary bias. *Journal of Informetrics*, *8*(2), 310–317. https://doi.org/10.1016/j.joi.2014.01.003

The random sample of five research evaluations that use Howard et al.'s counting method (Howard et al., 1987):

- Gardner, W. L., Lowe, K. B., Moss, T. W., Mahoney, K. T., & Cogliser, C. C. (2010). Scholarly leadership of the study of leadership: A review of The Leadership Quarterly's second decade, 2000–2009. *The Leadership Quarterly*, *21*(6), 922–958. https://doi.org/10.1016/j.leaqua.2010.10.003

- Tsai, C., & Lydia Wen, M. (2005). Research and trends in science education from 1998 to 2002: A content analysis of publication in selected journals. *International Journal of Science Education*, *27*(1), 3–14. https://doi.org/10.1080/0950069042000243727

The random sample of five research evaluations that use Sequence determines credit (Tscharntke et al., 2007):